\documentclass{article}
\usepackage{graphicx} \usepackage{amsmath,amssymb,mathtools}
\usepackage{booktabs}
\usepackage{amsthm}
\usepackage{placeins}
\usepackage{authblk}
\newtheorem{theorem}{Theorem}[section]
\newtheorem{lemma}[theorem]{Lemma}
\newtheorem{proposition}[theorem]{Proposition}

\newcommand{\dd}{\mathrm{d}}
\DeclareMathOperator{\sn}{sn}
\DeclareMathOperator{\dn}{dn}
\DeclareMathOperator{\sgn}{sgn}
\DeclareMathOperator{\sech}{sech}
\DeclareMathOperator{\csch}{csch}

\title{Semiclassical Averaging and Temporal Fluctuations of Quantum Dark Solitons \\
in the Lieb--Liniger Model}

\author[1]{Kota Kaneko
\thanks{\textit{Email address:} \texttt{m255003@edu.kaiyodai.ac.jp}}
}

\author[1]{Kohei Motegi\thanks{\textit{Email address:} \texttt{kmoteg0@kaiyodai.ac.jp}}}

\author[2]{Jun Sato
\thanks{\textit{Email address:} \texttt{j.sato@eng.t-kougei.ac.jp}}
}

\affil[1]{Faculty of Marine Technology, Tokyo University of Marine Science and Technology\\
Etchujima 2--1--6, Koto-Ku, Tokyo, 135-8533, Japan}
\affil[2]{
Faculty of Engineering, Tokyo Polytechnic University, Iiyama-minami 5-45-1,
Atsugi, Kanagawa, 243-0297, Japan
}
\date{\today}

\begin{document}

\maketitle

\begin{abstract}
We study the time evolution of quantum dark-soliton states of the Lieb–Liniger model, introduced by Sato–Kanamoto–Kaminishi–Deguchi as equal-weight superpositions of one-hole yrast states, and their relation to Gross–Pitaevskii (GP) solitons. We consider a simultaneous weak-coupling and large-system limit in which the repulsive coupling constant \(c\) tends to zero and the particle number \(N\) and ring circumference \(L\) tend to infinity, with the mean density \(\bar n=N/L\) and scaled circumference \(\Lambda=L\sqrt{c\bar n}\) fixed. We conjecture that, in this limit, the normalized quantum density and its evolution at fixed scaled time \(\tau=c\bar n t\) are described by a canonical-momentum average of normalized periodic one-soliton GP densities, each evolving at its own velocity. We explicitly parametrize the GP family by canonical momentum and provide numerical evidence for this correspondence. The different velocities explain the spreading of the averaged density notch despite shape preservation by individual solitons. In the infinite-line limit, we derive an analytic expression for the initial averaged profile, establish an inverse-fifth-power decay of its density deficit with distance from the soliton center, and express the Fourier transform of its deviation from the background as a series of Bickley–Naylor functions \(\operatorname{Ki}_3\). Following the spreading of the initial notch, we investigate temporal density fluctuations of the quantum dark solitons and propose a relation between long-time mode variances and second moments of GP Fourier coefficients, supported by numerical comparisons of individual mode variances and the total variance. Yrast transition frequencies distinguish the initial dephasing time scale from the longer scale associated with finite-time averaging of fluctuations.
\end{abstract}

\section{Introduction}\label{sec:intro}

The Lieb--Liniger model is a fundamental quantum integrable model of
one-dimensional bosons interacting through a contact potential. Its exact
solution describes the ground state and excitation spectrum, while the
Yang--Yang thermodynamic Bethe ansatz extends the description to thermal
equilibrium~\cite{LiebLiniger1963,Lieb1963II,YangYang1969,Gaudin1983}.
For repulsive interactions, the model interpolates between the weakly interacting regime and the impenetrable-boson (Tonks–Girardeau) limit
\cite{Girardeau1960}. Experiments on tightly confined ultracold atoms have
made it directly relevant to one-dimensional Bose gases, including the
observation of the Tonks--Girardeau regime and the quantum Newton's cradle
experiment~\cite{Kinoshita2004,Kinoshita2006}.

Developments in the quantum inverse scattering method have made it possible to study form factors and correlation functions of local operators of quantum integrable models, going beyond the energy spectrum
\cite{KBI1993,SlavnovBook}. Determinant formulas for Bethe-state norms,
scalar products, and field-operator matrix elements provide essential tools
for analytical calculations and accurate finite-size numerics
\cite{Korepin1982,Slavnov1989,KojimaKorepinSlavnov1997}.
For the Lieb--Liniger model, numerical form-factor summations have yielded
dynamical density and one-particle correlations over a wide range of couplings
\cite{CauxCalabrese2006,CauxCalabreseSlavnov2007,Caux2009,LiChengChenGuan2023},
and have also been used to study density correlations during nonequilibrium
evolution~\cite{GritsevRostunovDemler2010}. Substantial analytical results
have also been obtained for form factors and correlation functions
\cite{IIKS1990,Kitanine2012,Kozlowski2015,Pozsgay2011,KormosMussardoTrombettoni2009}.

Different weak-coupling regimes can be distinguished
in the Lieb--Liniger model, depending on how the
interaction strength scales with the particle number
and system size.
The ultraweak-coupling regime has been studied through
asymptotic solutions of the Bethe equations
~\cite{BatchelorGuanDunningLinks2005}.
In this work, we consider a simultaneous weak-coupling
and large-system limit, with the circumference in
healing-length units held fixed.
This limit is closely related to the
Gross--Pitaevskii (GP) equation
~\cite{Gross1961,Pitaevskii1961}.
The GP equation admits dark-soliton solutions
\cite{Tsuzuki1971,ZakharovShabat1973} and periodic nonlinear
waves~\cite{Tsuzuki1971}. Their relation to Lieb's type-II excitations
\cite{Lieb1963II,IshikawaTakayama1980} and to yrast states on a ring
\cite{KCU2009,KCU2010} provides a basis for studying quantum counterparts
of classical solitons. Here, a yrast state is the lowest-energy state at a
given total momentum.

Sato--Kanamoto--Kaminishi--Deguchi constructed localized quantum dark-soliton
states as equal-weight superpositions of one-hole yrast states
\cite{SatoPRL2012,SatoNJP2016}. At weak coupling, their initial density
profiles agree closely with those of periodic GP dark solitons, but the
initially localized density notch spreads under quantum time evolution. Related work connects
soliton profiles revealed by particle-position measurements with yrast
wave packets~\cite{Golletz2020}. The dynamics and scattering of multiple quantum dark
solitons have also been studied
\cite{IshiguroSatoEzakiNishinari2022,KinjoSatoDeguchi2023}.

In this paper, we investigate the original equal-weight one-hole states
in the weak-coupling semiclassical limit. 
We propose that the density profile is described by an
average of periodic one-soliton GP densities,
taken uniformly in normalized canonical momentum,
and give an explicit parametrization of the GP
family by this momentum.
The uniform measure is suggested by the
equal weights and equally spaced normalized momenta of the quantum yrast
states. 
In our numerical comparison, we keep the mean density
and the circumference in healing-length units fixed
while increasing the particle number and decreasing
the coupling.
The quantum density profiles become closer to the
GP average as the particle number increases.
At larger sizes, the GP average also gives a closer
approximation than a single periodic soliton fitted
to the minimum of the initial density.

Our proposal of averaging periodic GP soliton densities over the canonical momentum
also explains the spreading of the quantum density notch.
Each GP soliton preserves its dimensionless density profile as it moves, but
its velocity depends on the canonical momentum. Averaging these evolving
densities over the canonical momentum produces a notch that spreads with
time. We compare this averaged density with the quantum density during
time evolution using yrast state energies and density form factors, and find
numerical support for the correspondence.

We also analyze the infinite-line limit of the classical average.
In this limit, we obtain a closed expression for the
averaged initial density profile.
The density deficit from the background decays algebraically
as the inverse fifth power of distance, in contrast to the
exponential decay of an individual soliton.
The Fourier transform of the averaged density deviation
from the background is expressed as a series of
Bickley--Naylor functions~\cite{AbramowitzStegun1964}.
We also establish the corresponding limits of the periodic
averaged profile and Fourier coefficients.

Finally, we study temporal fluctuations of the quantum density after the initial notch has spread. Time averaging and dephasing provide a general framework for studying such fluctuations \cite{ShortFarrelly2012,deOliveira2018}. We express the long-time variance of each spatial Fourier mode in terms of the density form factors. We propose a semiclassical approximation of these variances using second moments of the Fourier coefficients over the same GP family, and test both the individual modes and their full sum numerically.
The yrast energy differences determine the frequencies within each spatial mode. The weighted spread of these frequencies characterizes initial dephasing, whereas the much smaller spacings between adjacent frequencies are associated with a longer time scale in finite-time averaging. Numerical comparisons of the initial-dephasing curves and the transition-frequency spacings support this separation of characteristic time scales.

This paper is organized as follows.
In Section~\ref{sec:GP-prelim}, we review the quantum
dark-soliton states and periodic GP solitons.
In Section~\ref{sec:quantum-to-gp-average}, we introduce
the canonical-momentum average of periodic GP solitons
and compare its initial density profile with the
quantum density profile and a fitted single GP soliton.
Section~\ref{sec:dynamic-gp-average} extends the
comparison between the quantum density and the GP average
to their time evolution.
In Section~\ref{sec:infinite-line-average}, we study
the infinite-line limit of the GP average and its
Fourier transforms.
Section~\ref{sec:fluctuation-relaxation} examines
temporal fluctuations of the quantum density,
their semiclassical GP predictions, and the
characteristic time scales of dephasing.
Section~\ref{sec:conclusion} contains our conclusions.

\section{
Quantum dark solitons of the Lieb-Liniger model
and periodic Gross--Pitaevskii solitons
}\label{sec:GP-prelim}
\label{sec:quantum-classical-setup}

\subsection{Lieb-Liniger model}\label{sec:LL-prelim}

We first review the Lieb--Liniger model
\cite{LiebLiniger1963,Gaudin1983}
and the construction of quantum dark-soliton states
\cite{SatoPRL2012,SatoNJP2016}.

In the units $2m=\hbar=1$, the repulsive
Lieb--Liniger Hamiltonian for $N$ bosons
on a ring of circumference $L$ is
\begin{equation}\label{eq:LL-H}
 H=-\sum_{j=1}^{N}\partial_{x_j}^2+2c\sum_{j<k}\delta(x_j-x_k),\qquad c>0.
\end{equation}

For the Bethe wave function to be an eigenfunction of the Hamiltonian with
periodic boundary conditions, the rapidities $k_j$, $j=1,\ldots,N$, must
satisfy the Bethe ansatz equations
\[
 e^{ik_jL}
 =
 \prod_{\substack{\ell=1\\ \ell\ne j}}^N
 \frac{k_j-k_\ell+ic}{k_j-k_\ell-ic},
 \qquad j=1,\ldots,N.
\]
For the repulsive case $c>0$, the Bethe rapidities
corresponding to physical eigenstates are real
and can be ordered as $k_1<\cdots<k_N$.
Taking the logarithm of the Bethe ansatz equations,
we obtain
\begin{equation*}
 Lk_j+\sum_{\ell\ne j}2\arctan\!\frac{k_j-k_\ell}{c}=2\pi I_j,
 \qquad j=1,\ldots,N.
\end{equation*}
Here $I_j$ are the distinct Bethe quantum numbers, which are integers for
odd $N$ and half-odd-integers for even $N$.  The momentum and energy of the
corresponding Bethe states are
\begin{equation*}
 P=\sum_{j=1}^Nk_j=\frac{2\pi}{L}\sum_{j=1}^NI_j,
 \qquad E=\sum_{j=1}^Nk_j^2.
\end{equation*}

We use the one-hole/type-II yrast family~\cite{Lieb1963II,SatoPRL2012,SatoNJP2016}
\begin{equation*}
 I_j^{(p)}=j-\frac{N+1}{2}+\mathbf 1_{\{j\ge N-p+1\}},
 \qquad p=0,\ldots,N-1.
\end{equation*}
Here \(\mathbf 1_{\{j\ge N-p+1\}}\) equals one if \(j\ge N-p+1\) and zero otherwise.
We denote the corresponding normalized eigenstate 
of the Lieb--Liniger Hamiltonian \eqref{eq:LL-H}
by $|P_p,N\rangle$.  Writing the rapidities of the $p$th state as $k_j^{(p)}$, we set
\begin{equation}\label{eq:onehole-PE}
 P_p:=\sum_{j=1}^N k_j^{(p)}=\frac{2\pi p}{L},\qquad
 E_p:=\sum_{j=1}^N\bigl(k_j^{(p)}\bigr)^2.
\end{equation}

Following \cite{SatoPRL2012}, we introduce
the equal-weight superposition of one-hole yrast states
\begin{equation}\label{eq:LL-PsiNstate}
 |\Psi_N\rangle
 =\frac1{\sqrt N}\sum_{p=0}^{N-1}|P_p,N\rangle.
\end{equation}
Its time evolution is
\begin{equation}\label{eq:LL-Xtime}
 |\Psi_N(t)\rangle
 =\frac1{\sqrt N}\sum_{p=0}^{N-1}e^{-iE_pt}|P_p,N\rangle.
\end{equation}
The momenta of the one-hole yrast states satisfy
\begin{equation}\label{eq:momentum-simple-match}
 \frac{P_p}{2\pi\bar n}=\frac{p}{N},\qquad \bar n=\frac NL.
\end{equation}
Thus the normalized momenta of the one-hole
yrast states are uniformly spaced.
In the semiclassical comparison below, the ratios
$p/N$ will be identified with the classical
normalized canonical momentum $s$
defined in \eqref{eq:proposal-sdef}.

We use the determinant representation of the density form factors
obtained by the quantum inverse scattering method
(algebraic Bethe ansatz)
\cite{SlavnovBook,KojimaKorepinSlavnov1997,
CauxCalabreseSlavnov2007}.

Let \(\hat\psi(x)\) and \(\hat\psi^\dagger(x)\) denote the bosonic annihilation and creation field operators, respectively, satisfying the canonical equal-time commutation relations
\[
\begin{aligned}
[\hat\psi(x),\hat\psi^\dagger(y)]
=\delta(x-y), \qquad
[\hat\psi(x),\hat\psi(y)]
=[\hat\psi^\dagger(x),\hat\psi^\dagger(y)]=0.
\end{aligned}
\]
The density operator is defined by \(\hat\rho(x)=\hat\psi^\dagger(x)\hat\psi(x)\).
For two distinct on-shell rapidity sets
\[
 \boldsymbol{k}=\{k_1,\ldots,k_N\},\qquad \boldsymbol{k}'=\{k'_1,\ldots,k'_N\},
\]
we denote the corresponding normalized Bethe eigenstates by
\(|\boldsymbol{k}\rangle\) and \(|\boldsymbol{k}'\rangle\).
Write \(P=\sum_jk_j\), \(P'=\sum_jk'_j\),
\(k_{j\ell}=k_j-k_\ell\), and \(k'_{j\ell}=k'_j-k'_\ell\).  Define
\begin{equation*}
 \widehat K(u)=\frac{2c}{u^2+c^2}.
\end{equation*}
The Gaudin matrix $G(\boldsymbol{k})$ is the $N \times N$ matrix
with matrix elements given by
\begin{equation*}
 G_{j \ell}(\boldsymbol{k})
 =\delta_{j\ell}\left[L+\sum_{a=1}^{N}\widehat K(k_j-k_a)\right]
 -\widehat K(k_j-k_\ell),
 \qquad j,\ell=1,\ldots,N.
\end{equation*}
This matrix is the Jacobian of the logarithmic Bethe equations
and appears in the Gaudin--Korepin norm formula
\cite{Gaudin1983,KBI1993,Korepin1982}.

The density form factor between normalized Bethe eigenstates,
used in our numerical calculations as in \cite{SatoPRL2012}, is given by
\cite{SlavnovBook,KojimaKorepinSlavnov1997,
CauxCalabreseSlavnov2007}
\begin{align}\label{eq:Sato-density-ff}
 \langle \boldsymbol{k}'|\hat\rho(0)|\boldsymbol{k} \rangle
 &=(-1)^{N(N+1)/2}(P-P')
 \left(\prod_{j,\ell=1}^{N}\frac1{k'_j-k_\ell}\right)\nonumber\\
 &\qquad\times\left(\prod_{j>\ell}k_{j\ell}k'_{j\ell}
 \sqrt{\frac{\widehat K(k'_{j\ell})}{\widehat K(k_{j\ell})}}\right)
 \frac{\det \mathcal U(\boldsymbol{k},\boldsymbol{k}')}{\sqrt{\det G(\boldsymbol{k})\det G(\boldsymbol{k}')}}.
\end{align}
Here \(\mathcal U(\boldsymbol{k},\boldsymbol{k}')\) is the \((N-1)\times(N-1)\) matrix with matrix elements given by
\begin{align*}
 \mathcal U_{j\ell}(\boldsymbol{k},\boldsymbol{k}')
 &=2\delta_{j\ell}\,\mathrm{Im}\!\left[
 \prod_{a=1}^{N}\frac{k'_a-k_j+ic}{k_a-k_j+ic}\right]\nonumber\\
 &\quad+
 \frac{\prod_{a=1}^{N}(k'_a-k_j)}
 {\prod_{a\ne j}^{N}(k_a-k_j)}
 \left[\widehat K(k_j-k_\ell)-\widehat K(k_N-k_\ell)\right],
\end{align*}
where $j,\ell=1,\ldots,N-1.$
For the yrast states we abbreviate
\begin{equation}\label{eq:Fpq-def}
 F_{p,q}:=\langle P_q,N|\hat\rho(0)|P_p,N\rangle.
\end{equation}
The diagonal matrix element is fixed by translation invariance,
\(F_{p,p}=N/L=\bar n\); \eqref{eq:Sato-density-ff} is used for \(p\ne q\).

For the equal-weight state \eqref{eq:LL-Xtime}, we introduce the 
density profile at time $t$
\begin{equation*}
 \rho_N(x,t):=\langle\Psi_N(t)|\hat\rho(x)|\Psi_N(t)\rangle.
\end{equation*}
Using \eqref{eq:LL-Xtime} and translation covariance of the density operator gives the exact form-factor expansion
\begin{equation}\label{eq:density-double-sum}
 \rho_N(x,t)
 =\frac1N\sum_{p,q=0}^{N-1}
 e^{i(P_p-P_q)x}e^{i(E_q-E_p)t}F_{p,q}.
\end{equation}
Let $x_0$ denote the position of the minimum of the initial density profile $\rho_N(x,0)$.  As observed in the numerical calculations \cite{SatoPRL2012,SatoNJP2016}, the initial density profile of the equal-weight state \eqref{eq:LL-PsiNstate} has its minimum at
\begin{equation}\label{eq:initial-notch-center}
 x_0=\frac{L}{2}.
\end{equation}

The density profile and its time evolution were studied numerically in
\cite{SatoPRL2012,SatoNJP2016}. At weak coupling, the initial profile agrees
closely with that of a periodic GP dark soliton, while the initially
localized quantum density notch spreads during time evolution (see also
\cite{KaminishiMoriMiyashita2020,SarviVianelloCaux2026} for other
superpositions). We next review the periodic GP solitons needed to formulate
our proposed description in terms of a canonical-momentum average of their
dimensionless densities.

\subsection{Gross--Pitaevskii equation}

The Gross--Pitaevskii (GP) equation~\cite{Gross1961,Pitaevskii1961} gives a
mean-field description of the weakly interacting Lieb--Liniger gas. Replacing
the field operator by a complex field in the Heisenberg equation with chemical
potential $\mu$ gives
\begin{equation}\label{eq:GP-dimensional}
 i\partial_t\psi=-\partial_x^2\psi+2c|\psi|^2\psi-\mu\psi.
\end{equation}
We choose $\mu=2c\bar n$ and use the dimensionless variables
\begin{equation}\label{eq:prelim-healing}
 \ell_c=(c\bar n)^{-1/2},\qquad
 y=\frac{x-x_0}{\ell_c},\qquad \tau=c\bar n\,t,\qquad
 \psi(x,t)=\sqrt{\bar n}\,\phi(y,\tau),
\end{equation}
where $x_0=L/2$ is the initial notch center in \eqref{eq:initial-notch-center}.
The dimensionless circumference is
\begin{equation}\label{eq:prelim-Lambda}
 \Lambda=\frac{L}{\ell_c}=N\sqrt{\gamma},\qquad \gamma=\frac{c}{\bar n}.
\end{equation}
Equation~\eqref{eq:GP-dimensional} then reads
\begin{equation*}
 i\partial_\tau\phi=-\partial_y^2\phi+2(|\phi|^2-1)\phi.
\end{equation*}

\paragraph{Periodic density profiles.}
We write a travelling wave as
$\phi(y,\tau)=\sqrt{R(z)}\,e^{i\Theta(z)-i\varpi\tau}$ with $z=y-u\tau$
and place the density minimum at $z=0$. We use the periodic one-soliton
solution of \cite[Eqs.~(40) and (46)]{SatoNJP2016}, with density roots
$R_j=a_j/\bar n$ and elliptic parameter $m_e=k^2$, where $k$ is the
elliptic modulus in that reference. Thus
\begin{equation}\label{eq:R-sn-main}
 \begin{aligned}
 R(z)&=R_1+(R_2-R_1)\sn^2(\kappa z\mid m_e)
       =R_3-\kappa^2\dn^2(\kappa z\mid m_e),\\
 \kappa^2&=R_3-R_1,\qquad m_e=\frac{R_2-R_1}{R_3-R_1}.
 \end{aligned}
\end{equation}
Here $0<R_1<R_2<R_3$ for a grey soliton, and $\sn$ and $\dn$ are the
Jacobi elliptic functions. Let
\begin{align}
 K=K(m_e)&:=\int_0^{\pi/2}\frac{\dd\alpha}{\sqrt{1-m_e\sin^2\alpha}},
 \label{eq:K-def}\\
 E=E(m_e)&:=\int_0^{\pi/2}\sqrt{1-m_e\sin^2\alpha}\,\dd\alpha.\notag
\end{align}
One notch per ring fixes
\begin{equation}\label{eq:kappa-def}
 \kappa=\frac{2K}{\Lambda}.
\end{equation}
The integral of $\dn^2(\kappa z\mid m_e)$ over one period is $2E/\kappa$.
The unit-mean condition $\int_0^\Lambda R(z)\,\dd z=\Lambda$ therefore gives
\begin{equation}\label{eq:R-explicit-main}
 R(z)=1+\frac{2\kappa E}{\Lambda}-\kappa^2\dn^2(\kappa z\mid m_e),
\end{equation}
with roots
\begin{equation}\label{eq:R-roots-main}
 R_3=1+\frac{2\kappa E}{\Lambda},\qquad
 R_1=R_3-\kappa^2,\qquad R_2=R_3-(1-m_e)\kappa^2.
\end{equation}
Appendix~\ref{app:GP-param-details} shows that $R_1\ge0$ precisely for
$0\le m_e\le m_*(\Lambda)$, where $m_*(\Lambda)$ is the unique zero of
$R_1$. The endpoints give uniform density and a black soliton, respectively.

\paragraph{Winding number and canonical momentum.}
For $R_1>0$, the phase derivative is
\begin{equation}\label{eq:proposal-current}
 \Theta'=\frac{u}{2}+\frac{J}{2R},\qquad J=\pm2\sqrt{R_1R_2R_3},
\end{equation}
which is the dimensionless form of the phase in
\cite[Eqs.~(30) and (41)]{SatoNJP2016}. The frequency is
$\varpi=R_1+R_2+R_3-2-u^2/4$. The periodic field requires the phase
increment $\Theta(\Lambda)-\Theta(0)$ to equal $2\pi w$ for some
$w\in\mathbb Z$. In terms of
\begin{equation}\label{eq:eta-def-main}
 \eta:=-\frac{R_2-R_1}{R_1}=-\frac{m_e\kappa^2}{R_1},
\end{equation}
define the complete elliptic integral of the third kind by
\begin{equation}\label{eq:Pi-complete-def}
 \Pi(\eta\mid m_e):=\int_0^{\pi/2}
 \frac{\dd\alpha}{(1-\eta\sin^2\alpha)\sqrt{1-m_e\sin^2\alpha}}.
\end{equation}
Integrating \eqref{eq:proposal-current} over one period gives
\begin{equation}\label{eq:GP-velocity-winding}
 2\pi w=\frac{u\Lambda+J I_R}{2},\qquad
 I_R:=\int_0^\Lambda\frac{\dd z}{R(z)}
     =\Lambda\frac{\Pi(\eta\mid m_e)}{K R_1}.
\end{equation}
The last equality follows by setting $\alpha=\operatorname{am}(\kappa z\mid m_e)$
in $I_R$ using
\(\eqref{eq:R-sn-main}\), where \(\operatorname{am}\) denotes the Jacobi amplitude.
For the canonical momentum
$P_{\rm GP,\Lambda}^{\rm per}=\bar n\int_0^\Lambda R\Theta'\,\dd z$,
the same phase derivative and unit mean give
\begin{equation}\label{eq:proposal-sdef}
 s:=\frac{P_{\rm GP,\Lambda}^{\rm per}}{2\pi\bar n}
   =\frac{\Lambda(u+J)}{4\pi}.
\end{equation}
Eliminating $u$ between \eqref{eq:GP-velocity-winding} and
\eqref{eq:proposal-sdef} yields the winding--momentum relation
\cite{KCU2009}
\begin{equation}\label{eq:F-Lambda-main}
 s=w-\sgn(J)F_\Lambda(m_e),\qquad
 F_\Lambda(m_e):=\frac{\Lambda\sqrt{R_1R_2R_3}}{2\pi}
       \left(\frac{\Pi(\eta\mid m_e)}{K R_1}-1\right).
\end{equation}
As proved in Appendix~\ref{app:GP-param-details}, $F_\Lambda$ increases
strictly from $0$ to $1/2$ on $[0,m_*(\Lambda)]$, with the value at the
black soliton defined by continuity.

The coordinate $s$ corresponds to the quantum labels $p/N$ in
\eqref{eq:momentum-simple-match}. The relation between periodic GP
solitons and the many-body yrast spectrum was studied in \cite{KCU2009,KCU2010}.
For the classical average we use $0\le s\le1$. The GP soliton family is parametrized by the following two branches:
\begin{equation}\label{eq:GP-momentum-choices}
 \begin{aligned}
 0<s<\tfrac12:&\quad w=0,\quad J<0,\quad F_\Lambda(m_e)=s,\\
 \tfrac12<s<1:&\quad w=1,\quad J>0,\quad F_\Lambda(m_e)=1-s.
 \end{aligned}
\end{equation}
The endpoints $s=0,1$ have uniform density, and $s=1/2$ is the black soliton.
For each $s$, the unique elliptic parameter is thus determined by
\begin{equation}\label{eq:GP-me-of-s}
 F_\Lambda(m_e(s))=\min\{s,1-s\}.
\end{equation}
For later use, define
\begin{equation}\label{eq:GP-A-Lambda-main}
 A_\Lambda(m_e):=
 2\sqrt{R_1R_2R_3}\,\frac{\Pi(\eta\mid m_e)}{K R_1}.
\end{equation}
By \eqref{eq:GP-velocity-winding}, the velocity is
\begin{equation}\label{eq:GP-velocity-s}
 u(s)=\begin{cases}
 A_\Lambda(m_e(s)), & 0<s<1/2,\\
 2\pi/\Lambda, & s=1/2,\\
 4\pi/\Lambda-A_\Lambda(m_e(s)), & 1/2<s<1.
 \end{cases}
\end{equation}
The middle value follows directly from $u=4\pi s/\Lambda-J$ in
\eqref{eq:proposal-sdef}, since $J\to0$ as $s\to1/2$. The pair $s,1-s$
has the same density profile and satisfies $u(s)+u(1-s)=4\pi/\Lambda$.

\section{From quantum dark solitons to an average of Gross--Pitaevskii solitons}
\label{sec:quantum-to-gp-average}

\subsection{
Canonical-momentum average and the conjectured
quantum--GP correspondence
}

We consider the limit $N\to\infty$ at fixed mean density $\bar n=N/L$ and fixed normalized circumference
$\displaystyle
 \Lambda=N\sqrt{\gamma}$, $\displaystyle \gamma=\frac{c}{\bar n}.
$
Equivalently,
\begin{equation}\label{eq:fixed-Lambda-scaling-expanded}
 L=\frac{N}{\bar n}\longrightarrow\infty,
 \qquad \gamma=\frac{\Lambda^2}{N^2}\longrightarrow0,
 \qquad c=\bar n\frac{\Lambda^2}{N^2}\longrightarrow0,
\end{equation}
while $L/\ell_c=\Lambda$ remains fixed. This is a weak-coupling semiclassical limit at fixed circumference in healing-length units.

For comparison with the GP densities, we define the dimensionless quantum density by rescaling \eqref{eq:density-double-sum}:
\begin{equation}\label{eq:RN-early-def}
 R_N(y,t):=\frac{\rho_N(x_0+\ell_c y,t)}{\bar n}.
\end{equation}
Here $y$ is the rescaled spatial coordinate \eqref{eq:prelim-healing}, whereas $t$ is the original time variable.

We first recall the comparison with a single GP soliton considered in \cite{SatoPRL2012,SatoNJP2016}.
At \(\tau=0\), we have \(z=y-u\tau=y\). We denote the dimensionless GP soliton density obtained by setting \(z=y\) in \eqref{eq:R-explicit-main} by \(R^{\mathrm{per}}_{m_e,\Lambda}(y)\).

\[
R^{\mathrm{per}}_{m_e,\Lambda}(y)
:=R(z)\big|_{z=y}
=1+\frac{2\kappa E(m_e)}{\Lambda}
-\kappa^2\operatorname{dn}^2(\kappa y\mid m_e).
\]

To make this comparison in the present normalization, we choose the elliptic parameter \(m_e=m^{\mathrm{fit}}_{e,N}\) so that the dimensionless GP density at \(y=0\) equals the initial dimensionless quantum density at the same point:

\begin{equation}\label{eq:single-GP-center-match}
 R_{m_{e,N}^{\rm fit},\Lambda}^{\rm per}(0)=R_N(0,0).
\end{equation}
The resulting profile $R_{m_{e,N}^{\rm fit},\Lambda}^{\rm per}(y)$ provides a single-soliton approximation.

For the limit  \eqref{eq:fixed-Lambda-scaling-expanded} with $\Lambda$ fixed, we instead consider an average of the initial dimensionless densities of periodic GP solitons. Using the normalized canonical momentum $s$ \eqref{eq:proposal-sdef}, we write
\begin{equation}\label{eq:R-s-Lambda-def}
 R_{s,\Lambda}^{\rm per}(y):=R_{m_e(s;\Lambda),\Lambda}^{\rm per}(y)
\end{equation}
where \(m_e(s;\Lambda)\) is the elliptic parameter determined by
\eqref{eq:GP-me-of-s}, with its dependence on \(\Lambda\) now displayed explicitly. We then define the canonical-momentum average by

\begin{equation}\label{eq:GP-average-main}
 R_{\mathrm{av},\Lambda}^{\rm per}(y):=\int_0^1 R_{s,\Lambda}^{\rm per}(y)\,\dd s.
\end{equation}
The uniform measure $\dd s$ is motivated by the equal weighting of the yrast states and the uniform spacing of their normalized momenta, $P_p/(2\pi\bar n)=p/N$. Once $\Lambda$ is fixed, this average contains no fitting parameter.

We conjecture that the initial dimensionless density of the Lieb-Liniger model  converges to this canonical-momentum average:
\begin{equation}\label{eq:fixed-Lambda-main-conjecture}
 R_N(y,0)\longrightarrow R_{\mathrm{av},\Lambda}^{\rm per}(y),
 \qquad N\to\infty\quad(\bar n,\Lambda\ \text{fixed}).
\end{equation}

\subsection{Numerical evidence of the conjecture at fixed $\Lambda$}
We compare the initial dimensionless density \(R_N(y,0)\) of the quantum dark-soliton state in the Lieb–Liniger model with the dimensionless density of the single GP soliton \(R^{\mathrm{per}}_{m^{\mathrm{fit}}_{e,N},\Lambda}(y)\) and the canonical-momentum average \(R^{\mathrm{per}}_{\mathrm{av},\Lambda}(y)\).
The numerical calculations are performed at
\begin{equation*}
 \Lambda=20,\qquad \bar n=1,\qquad L=N,\qquad c=\left(\frac{20}{N}\right)^2.
\end{equation*}
The initial dimensionless quantum density follows directly from the form-factor expansion in \eqref{eq:density-double-sum}. Setting $t=0$ and $x=L/2+\ell_c y$, dividing by $\bar n$, and grouping the terms by $m=|p-q|$, we obtain
\begin{equation*}
 R_N(y,0)=1+\frac{2}{N\bar n}\sum_{m=1}^{N-1}(-1)^m
 \left(\sum_{p=0}^{N-1-m}F_{p,p+m}\right)
 \cos\!\left(\frac{2\pi my}{\Lambda}\right).
\end{equation*}
The constant term follows from \(F_{p,p}=\bar n\). The factor \((-1)^m\) arises from expressing the spatial phase in the coordinate \(y=(x-L/2)/\ell_c\):
\[
e^{i(P_p-P_{p+m})x}
=e^{-i\pi m}e^{-2\pi imy/\Lambda}
=(-1)^m e^{-2\pi imy/\Lambda}.
\]
We use \eqref{eq:Sato-density-ff} for numerical evaluations
of the form factors.

Let $y_j$, $j=1,\ldots,M$, be the uniformly spaced sampling points in \([-\Lambda/2,\Lambda/2]\), with $M=40001$. For $R=
R^{\mathrm{per}}_{m^{\mathrm{fit}}_{e,N},\Lambda},
R_{\mathrm{av},\Lambda}^{\rm per}
$, we define the maximum and root-mean-square (RMS) errors on this grid by
\begin{equation*}
 E_{\max,N}[R]:=\max_{1\le j\le M}|R(y_j)-R_N(y_j,0)|,
\end{equation*}
and
\begin{equation*}
 E_{\mathrm{RMS},N}[R]:=\left[\frac1M\sum_{j=1}^M|R(y_j)-R_N(y_j,0)|^2\right]^{1/2}.
\end{equation*}

\begin{figure}[htbp]
\centering
\includegraphics[width=\linewidth]{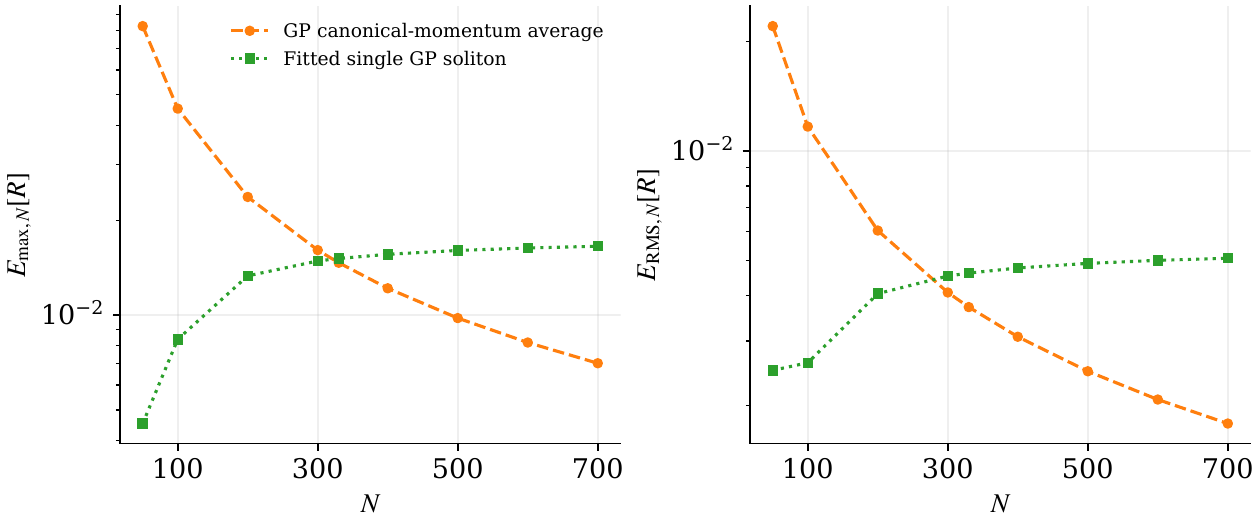}
\caption{Errors of the GP canonical-momentum average
$
R_{\mathrm{av},\Lambda}^{\rm per}(y)
$
and the fitted single GP soliton 
$
R^{\mathrm{per}}_{m^{\mathrm{fit}}_{e,N},\Lambda}(y)
$
relative to the initial quantum density profile $R_N(y,0)$ at fixed $\Lambda=20$, with $c=400/N^2$.
Left: $E_{\max,N}[R]$. Right: $E_{\mathrm{RMS},N}[R]$ for
$R=
R^{\mathrm{per}}_{m^{\mathrm{fit}}_{e,N},\Lambda},
R_{\mathrm{av},\Lambda}^{\rm per}
$.
Both vertical axes are logarithmic.}
\label{fig:fullsum-crossover-main}
\end{figure}

Figure~\ref{fig:fullsum-crossover-main} compares the two error measures. The average has a smaller maximum error than
the fitted single soliton for $N\ge330$ on the sampled sequence.
The RMS errors cross between $N=200$ and $N=300$, whereas the maximum
errors cross between $N=300$ and $N=330$. 
Thus, as the particle number increases along the sampled sequence, the canonical-momentum average 
$R_{\mathrm{av},\Lambda}^{\rm per}$
becomes more accurate than the fitted single GP soliton
$R_{m_{e,N}^{\rm fit},\Lambda}^{\rm per}$
according to both error measures.
These results support the fixed-$\Lambda$ conjecture in \eqref{eq:fixed-Lambda-main-conjecture}.

\begin{figure}[htbp]
\centering
\includegraphics[width=\linewidth]{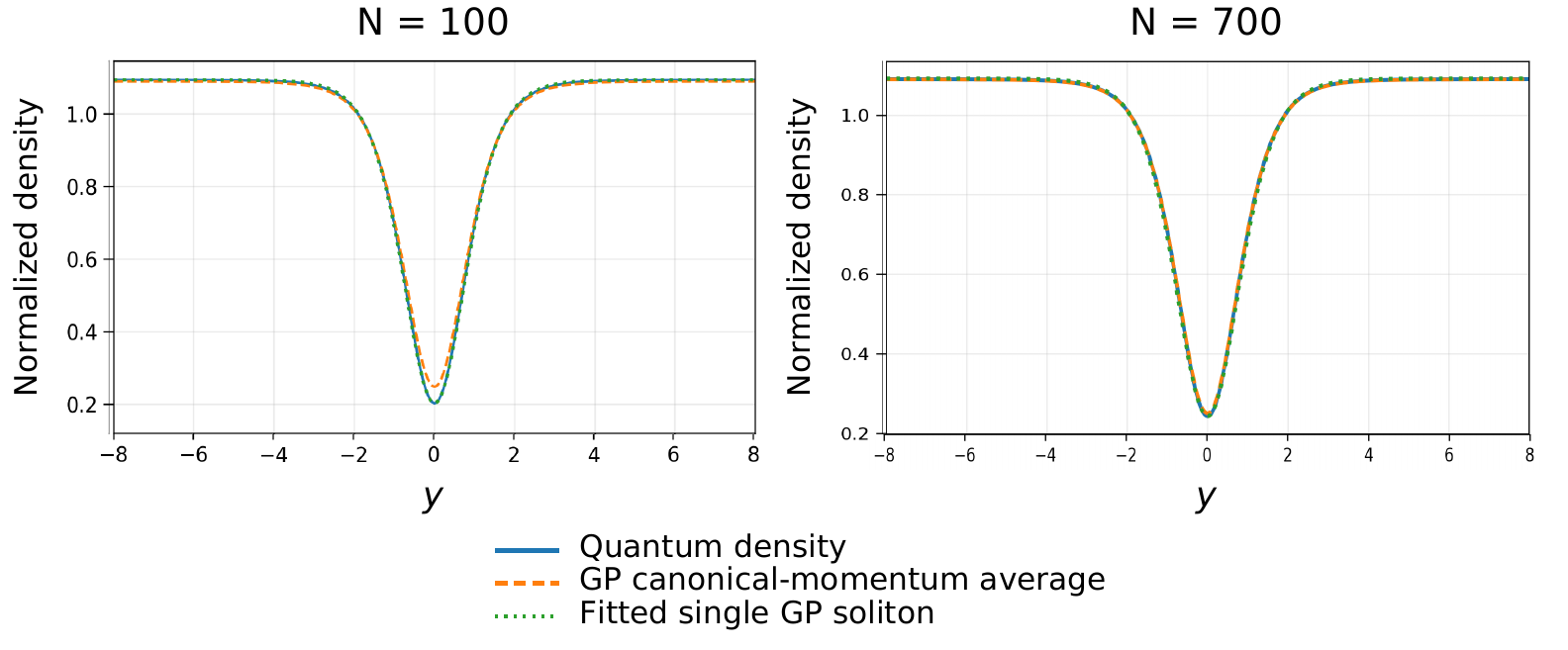}
\caption{Initial quantum density and the two classical profiles at fixed $\Lambda=20$, for $N=100$ (left) and $N=700$ (right).
Blue solid: $R_N(y,0)$. Orange dashed: the GP canonical-momentum average $R_{\mathrm{av},\Lambda}^{\rm per}(y)$. Green dotted: the fitted single GP soliton profile $R_{m_{e,N}^{\rm fit},\Lambda}^{\rm per}(y)$, with $m_{e,N}^{\rm fit}$ determined by \eqref{eq:single-GP-center-match}.
The coordinate is $y=(x-x_0)/\ell_c$, with $x_0=L/2$.}
\label{fig:fullsum-profile-comparison}
\end{figure}

\begin{figure}[htbp]
\centering
\includegraphics[width=.85\linewidth]{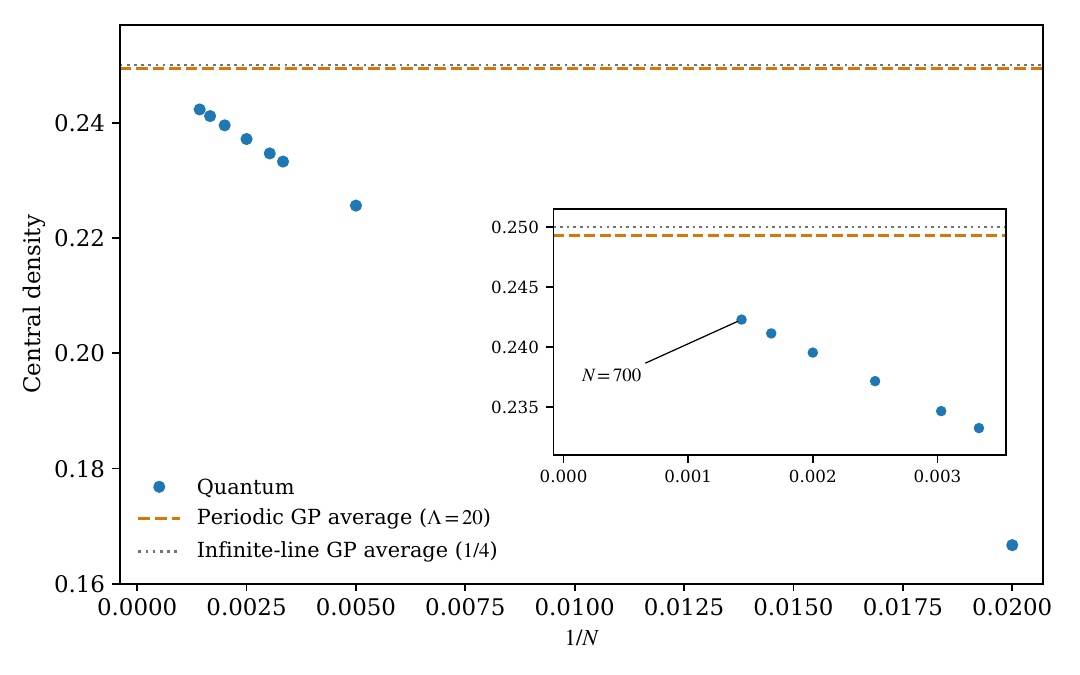}
\caption{Initial central density $R_N(0,0)$ of the Lieb-Liniger model versus $1/N$
at fixed $\Lambda=20$, with $\bar n=1$ and $c=400/N^2$,
for $N=50,100,200,300,330,400,500,600,700$.
The orange dashed line is the periodic GP canonical-momentum
average $R_{\mathrm{av},20}^{\mathrm{per}}(0)\simeq0.249$;
the grey dotted line is the infinite-line value $1/4$,
shown for reference.
The inset enlarges the large-$N$ region.}
\label{fig:central-density-convergence}
\end{figure}

Figure~\ref{fig:fullsum-profile-comparison} shows the initial quantum density profile \(R_N(y,0)\), the fitted single GP soliton profile \(R^{\mathrm{per}}_{m^{\mathrm{fit}}_{e,N},\Lambda}(y)\), and the canonical-momentum average \(R_{\mathrm{av},\Lambda}^{\mathrm{per}}(y)\) for \(N=100\) and \(N=700\) at fixed \(\Lambda=20\).
For $N=700$, the initial quantum central density is
$R_{700}(0,0)\simeq0.242$, compared with
$R_{\mathrm{av},20}^{\mathrm{per}}(0)\simeq0.249$
for the periodic GP canonical-momentum average at the same $\Lambda=20$.
Figure~\ref{fig:central-density-convergence} shows the approach
of the quantum central density to this fixed-$\Lambda$ comparison
value as $N$ increases.
For reference, the infinite-line value $R_{\mathrm{av}}(0)=1/4$,
derived in Section~\ref{sec:infinite-line-average}, is also shown;
it corresponds to the further limit $\Lambda\to\infty$.

\section{Time evolution of the periodic GP soliton average}
\label{sec:dynamic-gp-average}

In Section~\ref{sec:quantum-to-gp-average}, we compared the initial quantum density profile with the canonical-momentum average of periodic GP soliton densities. We now extend this comparison to time evolution. For each GP soliton, the normalized density \(|\phi(y,\tau)|^2=R(y-u\tau)\) retains its shape and moves with velocity \(u\). We average these normalized densities over the canonical momentum with the same uniform measure \(ds\), and compare the result with the time-dependent quantum density profile \(R_N(y,t)\).

\subsection{Time-dependent GP average and its Fourier representation}
\label{subsec:dyn-ensemble}

Recall the initial normalized GP soliton density
$R_{s,\Lambda}^{\rm per}(y)$ 
\eqref{eq:R-s-Lambda-def}.
The soliton with normalized canonical momentum $s$ moves with
dimensionless velocity $u(s)$ \eqref{eq:GP-velocity-s},
so its density at time $\tau=c\bar n t$ is
$R_{s,\Lambda}^{\rm per}(y-u(s)\tau)$.
Averaging these densities with respect to $s$, we define
\begin{equation}\label{eq:dyn-classical-density}
 R_{\mathrm{dyn},\Lambda}^{\mathrm{GP}}(y,\tau)
 :=\int_0^1 R_{s,\Lambda}^{\rm per}\bigl(y-u(s)\tau\bigr)\,\dd s.
\end{equation}
At $\tau=0$, this agrees with the initial
canonical-momentum average
$R_{\mathrm{av},\Lambda}^{\rm per}(y)$
 \eqref{eq:GP-average-main}.

We conjecture that, for fixed \(\bar n\), \(\Lambda\), and finite \(\tau\),
\[
R_N\!\left(y,\frac{\tau}{c\bar n}\right)
\longrightarrow R_{\mathrm{dyn},\Lambda}^{\mathrm{GP}}(y,\tau)
\qquad (N\to\infty).
\]

To compare the GP average with the quantum density, we express both densities as Fourier series with wave numbers $q_m=2\pi m/\Lambda$.
We first write the initial GP soliton density as
\begin{equation}\label{eq:gp-scaled-fourier-series-main}
 R_{s,\Lambda}^{\rm per}(y)
 =1+\frac2\Lambda\sum_{m=1}^{\infty}
 \mathfrak b_{m,\Lambda}^{\rm GP}(s)\cos(q_my),
\end{equation}
where
\begin{equation}\label{eq:gp-fourier-integral-main}
 \mathfrak b_{m,\Lambda}^{\rm GP}(s)
 :=\int_{-\Lambda/2}^{\Lambda/2}
 [R_{s,\Lambda}^{\rm per}(y)-1]e^{-iq_my}\,\dd y,
 \qquad m\ge1.
\end{equation}

The coefficients can be evaluated explicitly from  \eqref{eq:R-explicit-main}. Write $m_e=m_e(s;\Lambda)$, $K=K(m_e)$, and $E=E(m_e)$, and define the Jacobi nome by
\begin{equation*}
 \mathfrak q_e(s):=\exp\!\left[-\pi\frac{K(1-m_e)}{K(m_e)}\right].
\end{equation*}
The Fourier expansion for the Jacobi elliptic function \(\operatorname{dn}^2\)
\begin{equation*}
 \dn^2(v\mid m_e)=\frac EK+\frac{2\pi^2}{K^2}
 \sum_{m=1}^{\infty}\frac{m\mathfrak q_e(s)^m}{1-\mathfrak q_e(s)^{2m}}
 \cos\!\left(\frac{m\pi v}{K}\right)
\end{equation*}
and $\kappa=2K/\Lambda$ \eqref{eq:kappa-def} give
\begin{align*}
 R_{s,\Lambda}^{\rm per}(y)
 &=1+\frac{2\kappa E}{\Lambda}-\kappa^2\dn^2(\kappa y\mid m_e)\\
 &=1-\frac{8\pi^2}{\Lambda^2}\sum_{m=1}^{\infty}
 \frac{m\mathfrak q_e(s)^m}{1-\mathfrak q_e(s)^{2m}}\cos(q_my).
\end{align*}
Thus comparing with \eqref{eq:gp-scaled-fourier-series-main},
we have
\begin{equation}\label{eq:gp-scaled-fourier-main}
 \mathfrak b_{m,\Lambda}^{\rm GP}(s)
 =-\frac{4\pi^2m}{\Lambda}
 \frac{\mathfrak q_e(s)^m}{1-\mathfrak q_e(s)^{2m}},\qquad m\ge1.
\end{equation}

At time $\tau$, the density of this soliton is
\[
 R_{s,\Lambda}^{\rm per}\bigl(y-u(s)\tau\bigr)
 =1+\frac2\Lambda\operatorname{Re}\sum_{m=1}^{\infty}
 \mathfrak b_{m,\Lambda}^{\rm GP}(s)e^{-iq_mu(s)\tau}e^{iq_my}.
\]
Averaging over $s$ as in \eqref{eq:dyn-classical-density} therefore gives
\begin{equation}\label{eq:dyn-classical-fourier}
 R_{\mathrm{dyn},\Lambda}^{\mathrm{GP}}(y,\tau)
 =1+\frac2\Lambda\operatorname{Re}\sum_{m=1}^{\infty}
 B_{m,\Lambda}^{\rm dyn,GP}(\tau)e^{iq_my},
\end{equation}
where
\begin{equation*}
 B_{m,\Lambda}^{\rm dyn,GP}(\tau)
 :=\int_0^1\mathfrak b_{m,\Lambda}^{\rm GP}(s)
 e^{-iq_mu(s)\tau}\,\dd s.
\end{equation*}
In particular,
\begin{equation}\label{eq:A-first-moment}
 B_{m,\Lambda}^{\rm dyn,GP}(0)
 =\int_0^1\mathfrak b_{m,\Lambda}^{\rm GP}(s)\,\dd s.
\end{equation}

The exact quantum density has the corresponding expansion
\begin{equation}\label{eq:dyn-quantum-full}
 R_N\!\left(y,\frac{\tau}{c\bar n}\right)
 =1+\frac2\Lambda\operatorname{Re}\sum_{m=1}^{N-1}
 B_{m,N}^{\rm Q}(\tau)e^{iq_my},
\end{equation}
with
\begin{equation}\label{eq:dyn-quantum-B}
 B_{m,N}^{\rm Q}(\tau)
 :=\frac{\Lambda(-1)^m}{N\bar n}
 \sum_{p=0}^{N-1-m}F_{p,p+m}^{*}
 \exp\!\left[-i\frac{E_{p+m}-E_p}{c\bar n}\tau\right].
\end{equation}
Here a superscript $*$ denotes complex conjugation. This expansion follows from \eqref{eq:density-double-sum} by substituting $x=L/2+\ell_c y$ and $t=\tau/(c\bar n)$, and combining each pair of conjugate terms in the convention with spatial factor $e^{iq_my}$.

The solitons with canonical momenta $s$ and $1-s$ have the same
initial normalized density and velocities symmetric about $U=2\pi/\Lambda$:
\begin{equation}\label{eq:dyn-paired-velocities}
 R_{1-s,\Lambda}^{\rm per}(z)=R_{s,\Lambda}^{\rm per}(z),\qquad
 u(1-s)=2U-u(s).
\end{equation}
These relations follow from \eqref{eq:GP-velocity-s} and the equality
$m_e(s;\Lambda)=m_e(1-s;\Lambda)$
implied by \eqref{eq:GP-me-of-s}.
Setting $v(s)=u(s)-U$, splitting the integral in
\eqref{eq:dyn-classical-density} at $s=1/2$,
and changing variables $s\mapsto 1-s$ in the second
integral, we obtain from \eqref{eq:dyn-paired-velocities}
\[
 R_{\mathrm{dyn},\Lambda}^{\mathrm{GP}}(U\tau+z,\tau)
 =\int_0^{1/2}\!\left[
 R_{s,\Lambda}^{\rm per}(z-v(s)\tau)
 +R_{s,\Lambda}^{\rm per}(z+v(s)\tau)\right]\,\dd s.
\]
Since each initial density is even, we note from the expression above that the averaged density remains symmetric
about the moving point $y=U\tau$:
\begin{equation*}
 R_{\mathrm{dyn},\Lambda}^{\mathrm{GP}}(U\tau+z,\tau)
 =R_{\mathrm{dyn},\Lambda}^{\mathrm{GP}}(U\tau-z,\tau).
\end{equation*}
Its profile may nevertheless change with time.

The frequencies of the Lieb-Liniger model have a symmetry
corresponding to the GP solitons. In
\eqref{eq:dyn-quantum-B}, the state pair $(p,p+m)$ contributes the energy difference
\begin{equation}\label{eq:dyn-transition-frequency}
 \omega_{p,m,N}:=E_{p+m}-E_p,\qquad 0\le p\le N-1-m.
\end{equation}
The rapidities of yrast states \(p\) and \(N-p\) are related by
$k_j^{(N-p)}=-k_{N+1-j}^{(p)}+2\pi/L$ for $1\le p\le N-1$.
Together with $E_p=\sum_j(k_j^{(p)})^2$ and $P_p=2\pi p/L$,
we find
\begin{equation*}
 E_{N-p}=E_p-\frac{4\pi}{L}P_p+\frac{4\pi^2N}{L^2}
 =E_p+\frac{4\pi^2}{L^2}(N-2p).
\end{equation*}
Applying this identity to $p$ and $p+m$, and using
$L^2c\bar n=\Lambda^2$, we obtain
\begin{equation*}
 \frac{\omega_{p,m,N}+\omega_{N-m-p,m,N}}{c\bar n}
 =\frac{8\pi^2m}{\Lambda^2}=2q_mU,
 \qquad 1\le p\le N-m-1.
\end{equation*}
This is the Lieb–Liniger analogue of the GP frequency relation
$q_mu(s)+q_mu(1-s)=2q_mU$.

\subsection{Direct numerical comparison at fixed $\Lambda$}
\label{subsec:dyn-comparison}

We compare the quantum density profile \eqref{eq:dyn-quantum-full}
with the moving GP canonical-momentum average in
\eqref{eq:dyn-classical-density} and a fitted single periodic GP soliton.
The parameters we use are
\begin{equation*}
 \bar n=1,\qquad \Lambda=20,\qquad N=L=120,200,300,\qquad
 c=\left(\frac{20}{N}\right)^2.
\end{equation*}

The single GP soliton is chosen by matching the minimum of its initial density
profile to that of the quantum density profile, $R_N(0,0)$, as in \eqref{eq:single-GP-center-match}.
We then translate this fixed profile with each of its two velocities
allowed by the phase-winding condition. 
On the other hand,
the GP canonical-momentum average
has no fitted parameter.

\begin{figure}[htbp]
\centering
\includegraphics[width=.98\linewidth]{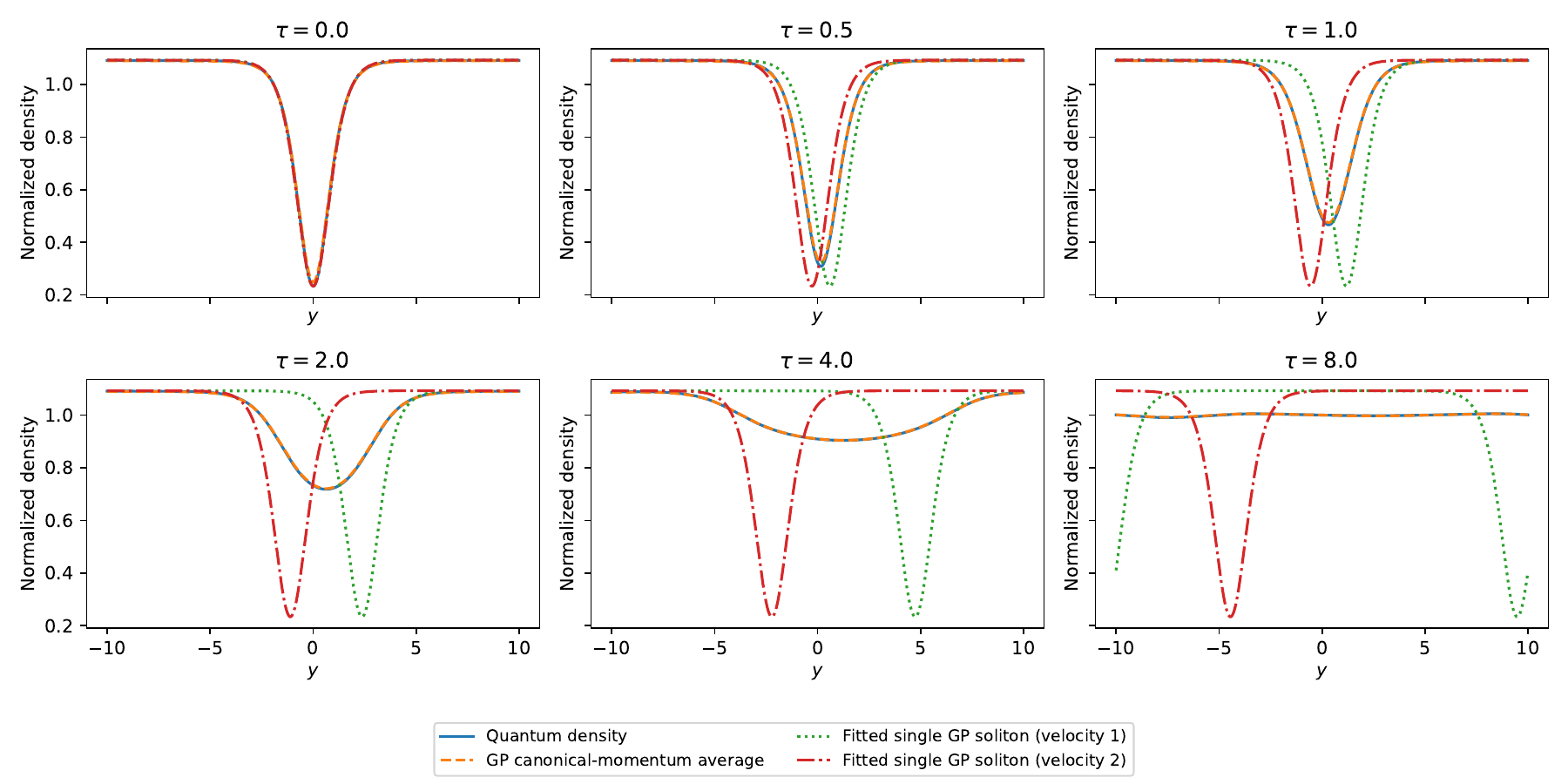}
\caption{The quantum density and the classical density profiles at
$N=300$ and $\Lambda=20$, for $\tau=0,0.5,1,2,4,8$.
The solid curve is the quantum density profile, the dashed curve is the moving
GP canonical-momentum average, and the dotted and dash-dotted curves
are the fitted single GP soliton translated with its two allowed velocities.}
\label{fig:dyn-N300}
\end{figure}

Figure~\ref{fig:dyn-N300} compares the density profiles at six times.
The quantum density changes shape as its initial notch spreads, whereas
each single GP soliton retains its shape. The moving GP average reproduces
the time evolution of the quantum density profile much more closely.

\begin{figure}[htbp]
\centering
\includegraphics[width=\linewidth]{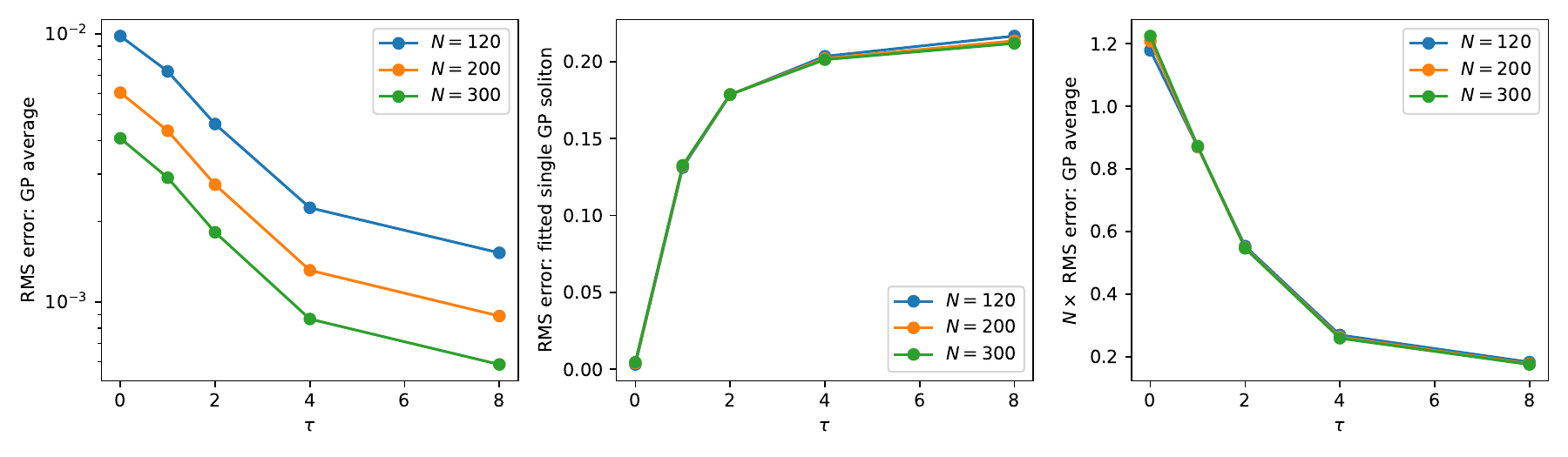}
\caption{Spatial RMS errors at fixed $\Lambda=20$ for $N=120,200,300$.
Left: the moving GP canonical-momentum average (logarithmic vertical scale).
Middle: the smaller of the errors obtained from the two allowed single-soliton velocities.
Right: the GP-average error multiplied by $N$.
The sampled times $\tau=0,1,2,4,8$ are connected by lines.}
\label{fig:dyn-rms}
\end{figure}

We measure the spatial RMS difference over one period using 1280 equally
spaced points. Figure~\ref{fig:dyn-rms} compares the error of the GP average
with the smaller of the errors obtained from the two allowed single-soliton
velocities. At the sampled positive times, the GP average has a substantially
smaller error for \(N=120,200,300\). Its error decreases with $N$,
and the curves nearly collapse when multiplied by $N$.
This suggests an
approximately $N^{-1}$ finite-size correction over this range.

\section{Infinite-line limit of the GP canonical-momentum average}
\label{sec:infinite-line-average}

We consider the infinite-line limit $\Lambda\to\infty$
of the periodic GP one-soliton family, keeping $y$ fixed.
We first obtain the limiting soliton profile
and then calculate its canonical-momentum averages,
including the time-dependent density profile and Fourier transforms.

\subsection{Single-soliton degeneration and convergence of the static average}

Recall that $K(m_e)$ is the complete elliptic integral of the first kind
defined by \eqref{eq:K-def}.  For fixed $0<\kappa<1$ and sufficiently
large $\Lambda$, the periodicity condition \eqref{eq:kappa-def} determines
the elliptic parameter $m_e=m_e(\kappa;\Lambda)$ through
\begin{equation}\label{eq:decompact-K-condition}
 K(m_e)=\frac{\kappa\Lambda}{2}.
\end{equation}
As $\Lambda\to\infty$, we have $m_e\uparrow1$, and the standard limits of
elliptic functions give
\begin{equation}\label{eq:decompact-elliptic-limits}
 E(m_e)\longrightarrow1,\qquad
 \dn(u\mid m_e)\longrightarrow\sech u,
\end{equation}
where $\sech u=1/\cosh u$.  From \eqref{eq:R-roots-main} and \eqref{eq:eta-def-main}, we obtain
\begin{equation}\label{eq:root-limits-main}
 R_1\longrightarrow1-\kappa^2,\qquad
 R_2\longrightarrow1,\qquad
 R_3\longrightarrow1,\qquad
 \eta\longrightarrow\eta_0:=-\frac{\kappa^2}{1-\kappa^2}<0.
\end{equation}
The periodic density \eqref{eq:R-explicit-main} therefore converges to
\begin{equation}\label{eq:grey-density-main}
 R_\xi(y)=1-(1-\xi^2)
 \sech^2\!\left(\sqrt{1-\xi^2}\,y\right),
\end{equation}
where
\begin{equation}\label{eq:xi-signed-main}
 \xi:=-\sgn(J)\sqrt{1-\kappa^2},\qquad -1\le\xi\le1.
\end{equation}

We next determine the relation between \(s\) and \(\xi\) in the infinite-line limit.
First, recall the definition of $F_\Lambda$ in \eqref{eq:F-Lambda-main}.  The complete elliptic integrals
$K(m_e)$
\eqref{eq:K-def}
and $\Pi(\eta\mid m_e)$
\eqref{eq:Pi-complete-def} both diverge
as $m_e\uparrow1$.  Their divergent parts cancel after subtracting
$K(m_e)/(1-\eta)$, because
\[
 \frac{1}{1-\eta\sin^2\varphi}-\frac{1}{1-\eta}
 =-\frac{\eta\cos^2\varphi}
 {(1-\eta)(1-\eta\sin^2\varphi)}.
\]
Indeed, the subtraction gives
\begin{align*}
 \Pi(\eta\mid m_e)-\frac{K(m_e)}{1-\eta}
 &=-\frac{\eta}{1-\eta}
 \int_0^{\pi/2}
 \frac{\cos^2\varphi\,\dd\varphi}
 {(1-\eta\sin^2\varphi)
  \sqrt{1-m_e\sin^2\varphi}}.
\end{align*}
The factor $\cos^2\varphi$ removes the endpoint singularity in the
limit.  Using $\eta\to\eta_0<0$ from \eqref{eq:root-limits-main}
and the substitution $t=\sin\varphi$, we obtain
\begin{align}\label{eq:Pi-renorm-limit-main}
 \Pi(\eta\mid m_e)-\frac{K(m_e)}{1-\eta}
 &\longrightarrow -\frac{\eta_0}{1-\eta_0}
 \int_0^1\frac{\dd t}{1-\eta_0t^2}
 =\frac{\sqrt{-\eta_0}}{1-\eta_0}
 \arctan\sqrt{-\eta_0}.
\end{align}
Using $R_2=R_1(1-\eta)$ and \eqref{eq:decompact-K-condition},
we rewrite \eqref{eq:F-Lambda-main} as
\[
\begin{aligned}
 F_\Lambda(m_e)
 ={}&\frac{\Lambda\sqrt{R_1R_2R_3}}{2\pi}
 \left(\frac1{R_2}-1\right)
 +\frac{\sqrt{R_1R_2R_3}}{\pi\kappa R_1}
 \left[\Pi(\eta\mid m_e)-\frac{K(m_e)}{1-\eta}\right].
\end{aligned}
\]
By \eqref{eq:R-roots-main}, the first term equals
\[
 \frac{\sqrt{R_1R_2R_3}}{2\pi R_2}
 \left[\Lambda\kappa^2(1-m_e)-2\kappa E(m_e)\right].
\]
The standard asymptotic formula
$K(m_e)=\frac12\log\!\bigl(16/(1-m_e)\bigr)+o(1)$ as $m_e\uparrow1$,
together with \eqref{eq:decompact-K-condition}, gives
$1-m_e\sim16e^{-\kappa\Lambda}$, and hence
$\Lambda(1-m_e)\to0$.  Thus, by
\eqref{eq:decompact-elliptic-limits} and \eqref{eq:root-limits-main},
the first term tends to $-\kappa\sqrt{1-\kappa^2}/\pi$.
For the second term, \eqref{eq:Pi-renorm-limit-main} and
$\arctan\sqrt{-\eta_0}=\arcsin\kappa$ give
\[
 \frac{\sqrt{R_1R_2R_3}}{\pi\kappa R_1}
 \left[\Pi(\eta\mid m_e)-\frac{K(m_e)}{1-\eta}\right]
 \longrightarrow \frac{\arcsin\kappa}{\pi}.
\]
Combining the two contributions, we obtain
\begin{equation}\label{eq:F-infinite-kappa}
 F_\Lambda(m_e(\kappa;\Lambda))\longrightarrow
 \frac1\pi\left[\arcsin\kappa-\kappa\sqrt{1-\kappa^2}\right].
\end{equation}
From the momentum choices \eqref{eq:GP-momentum-choices} and the sign
convention \eqref{eq:xi-signed-main}, we note that $\xi>0$ corresponds
to $J<0$ and $s=F_\Lambda(m_e)$, whereas $\xi<0$ corresponds to
$J>0$ and $s=1-F_\Lambda(m_e)$.  Using \eqref{eq:F-infinite-kappa} and
$\kappa=\sqrt{1-\xi^2}$, the two cases give
\begin{equation}\label{eq:s-xi-infinite}
 s(\xi)=\frac1\pi\left[
  \arccos\xi-\xi\sqrt{1-\xi^2}\right],
 \qquad -1\le\xi\le1.
\end{equation}
The map is continuous and strictly decreasing from $s(-1)=1$ to
$s(1)=0$, with
\begin{equation*}
 \left|\frac{\dd s}{\dd\xi}\right|
 =\frac2\pi\sqrt{1-\xi^2}.
\end{equation*}
Thus an average over uniformly distributed canonical momenta becomes
\begin{equation}\label{eq:measure-change-main}
 \int_0^1 f(\xi(s))\,\dd s
 =\frac2\pi\int_{-1}^1 f(\xi)\sqrt{1-\xi^2}\,\dd\xi.
\end{equation}
Applying \eqref{eq:measure-change-main} to \eqref{eq:grey-density-main}
produces the static limiting profile:
\begin{align}\label{eq:Rav-main}
 R_{\rm av}(y)
 &:=\int_0^1R_{\xi(s)}(y)\,\dd s
 =1-\frac2\pi\int_{-1}^1
 (1-\xi^2)^{3/2}
 \sech^2\!\left(\sqrt{1-\xi^2}\,y\right)\dd\xi.
\end{align}
The same static infinite-line average is implicit in the weak-coupling
GP approximation in Sarvi-Vianello-Caux
\cite{SarviVianelloCaux2026}.

From  \eqref{eq:Rav-main}, we have
\begin{equation*}
 R_{\rm av}(0)
 =1-\frac2\pi\int_{-1}^1(1-\xi^2)^{3/2}\,\dd\xi
 =\frac14.
\end{equation*}

Unlike each individual grey soliton, the average has an algebraic tail.
Substituting $\kappa=\sqrt{1-\xi^2}$ into \eqref{eq:Rav-main} gives
\begin{equation*}
 R_{\rm av}(y)
 =1-\frac4\pi\int_0^1
 \frac{\kappa^4}{\sqrt{1-\kappa^2}}\sech^2(\kappa y)\,\dd\kappa.
\end{equation*}
For $y\ne0$, the substitution $u=\kappa|y|$ gives
\[
 R_{\rm av}(y)
 =1-\frac4{\pi|y|^5}\int_0^{|y|}
 \frac{u^4\sech^2u}{\sqrt{1-u^2/y^2}}\,\dd u.
\]
As $|y|\to\infty$, the integrand is dominated by a constant times
$u^4\sech^2u$ on $0\le u\le|y|/2$; on the remaining interval,
the exponential decay of $\sech^2u$ controls the integrable
square-root singularity at $u=|y|$.  Consequently,
\begin{equation*}
 \begin{aligned}
 R_{\rm av}(y)
 &=1-\frac4{\pi|y|^5}\int_0^\infty u^4\sech^2u\,\dd u
   +o(|y|^{-5})
 =1-\frac{7\pi^3}{60|y|^5}+o(|y|^{-5}).
 \end{aligned}
\end{equation*}

\begin{figure}[htbp]
\centering
\includegraphics[width=.72\linewidth]{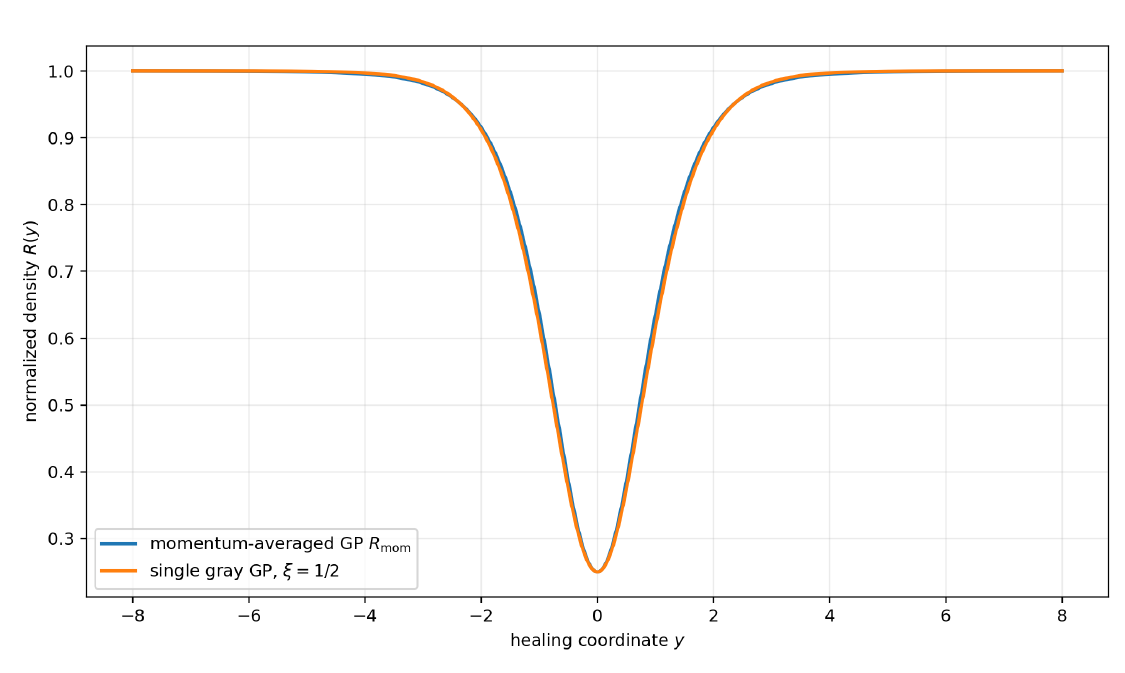}
\caption{Comparison of the infinite-line canonical-momentum average
$R_{\rm av}(y)$ \eqref{eq:Rav-main} and the single grey soliton
$R_\xi(y)$ \eqref{eq:grey-density-main} with $\xi=1/2$.
The profiles have the same central density but differ in curvature
and tail behavior.}
\label{fig:average-vs-single}
\end{figure}

To justify the infinite-line limit of the periodic canonical-momentum
average, we establish the following theorem.

\begin{theorem}
\label{thm:classical-decompact-main}
For every integer $k\ge0$ and every $Y<\infty$,
\begin{equation}\label{eq:classical-profile-decompact-main}
 \sup_{|y|\le Y}\left|
 \partial_y^kR_{\mathrm{av},\Lambda}^{\rm per}(y)
 -\partial_y^kR_{\rm av}(y)\right|\longrightarrow0
 \qquad(\Lambda\to\infty).
\end{equation}
\end{theorem}

The proof is given in Appendix B.1.

\subsection{Time-dependent infinite-line average}
\label{subsec:dyn-infinite-line}

We now take the infinite-line limit of the time-dependent GP average
\eqref{eq:dyn-classical-density}.  Using $R_2=R_1(1-\eta)$, we have
\[
 \begin{aligned}
 \frac{\Pi(\eta\mid m_e)}{K(m_e)R_1}
 &=\frac{1}{R_2}
 +\frac{\Pi(\eta\mid m_e)-K(m_e)/(1-\eta)}{K(m_e)R_1}
 \longrightarrow1.
 \end{aligned}
\]
Indeed, the numerator of the second term has a finite limit by
\eqref{eq:Pi-renorm-limit-main}, while $K(m_e)\to\infty$ and
$R_1\to1-\kappa^2>0$; moreover $R_2\to1$ by
\eqref{eq:root-limits-main}.  Consequently, the limit of
$A_\Lambda(m_e(s))$
\eqref{eq:GP-A-Lambda-main} becomes
\[
 A_\Lambda(m_e(s))\longrightarrow
 2\sqrt{1-\kappa^2}=2|\xi(s)|.
\]
By \eqref{eq:GP-velocity-s} and \eqref{eq:xi-signed-main},
we have $u(s)=A_\Lambda(m_e(s))$ and $\xi(s)>0$
for $s<1/2$, whereas
$u(s)=4\pi/\Lambda-A_\Lambda(m_e(s))$ and $\xi(s)<0$
for $s>1/2$.
Since $4\pi/\Lambda\to0$, both cases yield
\[
u(s)\longrightarrow 2\xi(s)
\]
for each fixed $s\in(0,1)$ with $s\ne1/2$.
Combining this velocity limit with the single-soliton profile
\eqref{eq:grey-density-main} and the canonical-momentum measure
\eqref{eq:measure-change-main}, we obtain
\begin{equation}\label{eq:dyn-infinite-profile}
 \begin{aligned}
 R_{\rm dyn,\infty}(y,\tau)
 &=\int_0^1 R_{\xi(s)}\!\left(y-2\xi(s)\tau\right)\,\dd s\\
 &=1-\frac2\pi\int_{-1}^1(1-\xi^2)^{3/2}
 \sech^2\!\left(\sqrt{1-\xi^2}(y-2\xi\tau)\right)\,\dd\xi.
 \end{aligned}
\end{equation}
At $\tau=0$, this reduces to $R_{\rm av}(y)$.

The convergence of the finite-ring GP average is established by the following proposition.

\begin{proposition}
\label{prop:dyn-decompact-uniform}
For all finite $Y,T>0$,
\begin{equation}\label{eq:dyn-decompact-uniform}
 \sup_{\substack{|y|\le Y\\|\tau|\le T}}
 \left|R_{\mathrm{dyn},\Lambda}^{\rm GP}(y,\tau)
       -R_{\rm dyn,\infty}(y,\tau)\right|\longrightarrow0.
\end{equation}
\end{proposition}
The proof is in Appendix B.1.

\begin{proposition}
\label{prop:dyn-infinite-filling}
The profile \eqref{eq:dyn-infinite-profile} satisfies
\begin{align}
 R_{\rm dyn,\infty}(0,\tau)
 &=\frac14+\frac5{16}\tau^2+O(\tau^4)
 \qquad(\tau\to0),\label{eq:dyn-infinite-short}\\
 R_{\rm dyn,\infty}(y,\tau)
 &=1-\frac2{\pi\tau}+o(\tau^{-1})
 \qquad(\tau\to+\infty),\quad y\ \text{fixed}.
 \label{eq:dyn-infinite-long}
\end{align}
\end{proposition}
\begin{proof}
At $y=0$, insert
$\sech^2 z=1-z^2+O(z^4)$ with
$z=-2\xi\sqrt{1-\xi^2}\,\tau$ into the integrand of
\eqref{eq:dyn-infinite-profile}.  The required integrals are
\[
 \frac2\pi\int_{-1}^1(1-\xi^2)^{3/2}\,\dd\xi=\frac34,
 \qquad
 \frac8\pi\int_{-1}^1\xi^2(1-\xi^2)^{5/2}\,\dd\xi
 =\frac5{16}.
\]
This proves \eqref{eq:dyn-infinite-short}.

To prove \eqref{eq:dyn-infinite-long}, fix $y$ and let
$\tau\to\infty$.
The leading contribution to
\eqref{eq:dyn-infinite-profile} comes from $\xi=O(\tau^{-1})$.
Introducing $v=2\xi\tau-y$, we have
$\xi=(v+y)/(2\tau)$ and $\dd\xi=\dd v/(2\tau)$.
Thus,
\[
\begin{aligned}
R_{\rm dyn,\infty}(y,\tau)
&=1-\frac1{\pi\tau}
\int_{-2\tau-y}^{2\tau-y}
\left(1-\frac{(v+y)^2}{4\tau^2}\right)^{3/2}\\
&\quad\times\sech^2\!\left(
v\sqrt{1-\frac{(v+y)^2}{4\tau^2}}
\right)\,\dd v.
\end{aligned}
\]
For each fixed \(v\), the integrand converges to \(\operatorname{sech}^2 v\) as \(\tau\to\infty\). Since \(\int_{-\infty}^{\infty}\operatorname{sech}^2v\,dv=2\), this formally gives
$R_{\rm dyn,\infty}(y,\tau)=1-2/(\pi\tau)+o(\tau^{-1})$.

To justify taking the limit under the integral, divide the
integration domain into $|\xi|\le\delta$ and $|\xi|>\delta$,
with fixed $0<\delta<1$. One can show that
\[
 \begin{aligned}
 &\frac2\pi\int_{|\xi|\le\delta}(1-\xi^2)^{3/2}
 \sech^2\!\left(\sqrt{1-\xi^2}(y-2\xi\tau)\right)\,\dd\xi\\
 &\quad=\frac1{\pi\tau}
 \left[\int_{-\infty}^{\infty}\sech^2v\,\dd v+o(1)\right],
 \end{aligned}
\]
while
\[
 \begin{aligned}
 &\int_{\delta<|\xi|\le1}(1-\xi^2)^{3/2}
 \sech^2\!\left(\sqrt{1-\xi^2}(y-2\xi\tau)\right)\,\dd\xi
 =O(\tau^{-5}).
 \end{aligned}
\]
These estimates justify the asymptotic formula
\eqref{eq:dyn-infinite-long}.
\end{proof}

\begin{figure}[htbp]
\centering
\includegraphics[width=.80\linewidth]{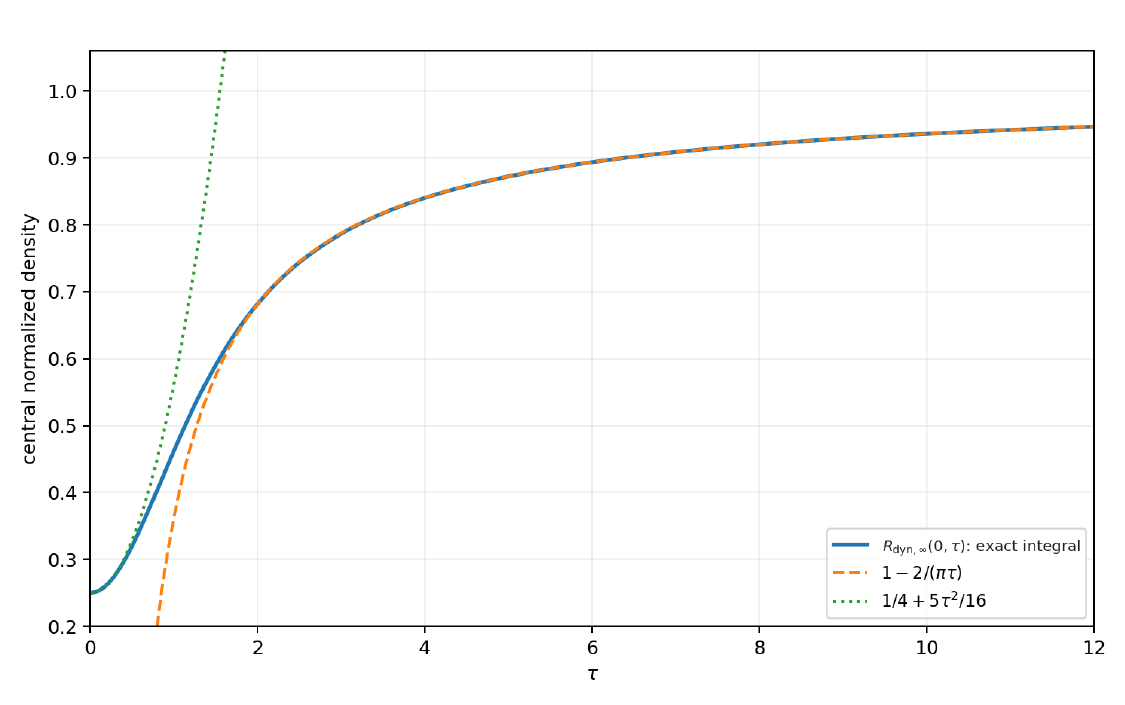}
\caption{
Time evolution of $R_{\rm dyn,\infty}(0,\tau)$ for the moving GP
canonical-momentum average in the infinite-line limit,
compared with the short-time expansion
\eqref{eq:dyn-infinite-short} and the long-time asymptotic
formula \eqref{eq:dyn-infinite-long}.
\label{fig:dyn-infinite-asymptotics}}
\end{figure}

The preceding results concern the classical GP ensemble. Assuming
 the quantum--GP correspondence proposed in
 Sections~\ref{sec:quantum-to-gp-average} and
 \ref{sec:dynamic-gp-average}, we expect the limit
\[
 \lim_{\Lambda\to\infty}
 \left[\lim_{\substack{N\to\infty\\\bar n,\,\Lambda\ \mathrm{fixed}}}
 R_N\!\left(y,\frac{\tau}{c\bar n}\right)\right]
 =R_{\rm dyn,\infty}(y,\tau)
\]
for fixed $y$ and $\tau$. The quantum--GP correspondence remains
conjectural.

\subsection{Fourier transforms and the infinite-line limit}

We now consider the Fourier transforms of the limiting soliton profiles.

From the single-soliton profile \eqref{eq:grey-density-main}, the Fourier
transform of $R_\xi(y)-1$ is, for $r>0$,
\begin{align}\label{eq:Bxi-main}
 B_\xi(r)
 &:=\int_{\mathbb R}[R_\xi(y)-1]e^{-2\pi i r y}\,\dd y
 =-2\pi^2r\,\csch\!\left(
   \frac{\pi^2r}{\sqrt{1-\xi^2}}\right).
\end{align}
Here $\csch z=1/\sinh z$, and $B_\xi(0)$ is defined by continuity
for $|\xi|<1$.

Using the canonical-momentum measure \eqref{eq:measure-change-main},
we obtain
\begin{equation}\label{eq:Bav-integral-main}
 B_{\rm av}(r):=\int_0^1B_{\xi(s)}(r)\,\dd s
 =-4\pi r\int_{-1}^1\sqrt{1-\xi^2}\,
 \csch\!\left(\frac{\pi^2r}{\sqrt{1-\xi^2}}\right)\dd\xi.
\end{equation}
Expanding $\csch z=2\sum_{k=0}^\infty e^{-(2k+1)z}$ and making the
substitutions $\xi=\sin\theta$ and $\sec\theta=\cosh t$, we find
\begin{equation}\label{eq:Bav-Ki-main}
 B_{\rm av}(r)=-16\pi r\sum_{k=0}^\infty
 \operatorname{Ki}_3\!\left((2k+1)\pi^2r\right),
\end{equation}
where the Bickley--Naylor function is defined by
\cite[(11.2.10)]{AbramowitzStegun1964} as
\begin{equation}\label{eq:Ki-def-main}
 \operatorname{Ki}_\nu(z)
 :=\int_0^\infty e^{-z\cosh t}\sech^\nu t\,\dd t.
\end{equation}

We now relate \(B_{\rm av}(r)\) to the periodic GP Fourier coefficients. The corresponding scaled Fourier mode is
\begin{equation}\label{eq:r-healing-main}
 r:=\frac m\Lambda.
\end{equation}
Using the Jacobi nome
$\mathfrak q_e(s)=\exp[-\pi K(1-m_e)/K(m_e)]$ and the periodicity
condition $K(m_e)=\kappa\Lambda/2$, the periodic Fourier coefficient
\eqref{eq:gp-scaled-fourier-main} can be rewritten as
\begin{equation}\label{eq:gp-csch-fourier-main}
 \mathfrak b_{m,\Lambda}^{\rm GP}(s)
 =-\frac{2\pi^2m}{\Lambda}
 \csch\!\left(\frac{2\pi mK(1-m_e)}{\kappa\Lambda}\right).
\end{equation}
Taking the infinite-line limit of the elliptic parameters and averaging
over the canonical momentum as in \eqref{eq:A-first-moment}, we recover
the averaged Fourier transform \eqref{eq:Bav-integral-main}:
\begin{equation*}
 \begin{aligned}
 B_{m_\Lambda,\Lambda}^{\rm dyn,GP}(0)
 &=\int_0^1\mathfrak b_{m_\Lambda,\Lambda}^{\rm GP}(s)\,\dd s
 \longrightarrow B_{\rm av}(r),\qquad
 \frac{m_\Lambda}{\Lambda}\longrightarrow r>0
 \quad(\Lambda\to\infty).
 \end{aligned}
\end{equation*}
The convergence is locally uniform for $r>0$.
Weighted \(L^1\) convergence is also established in Appendix B.2.

\section{Temporal fluctuations and dephasing}
\label{sec:fluctuation-relaxation}

Sections~\ref{sec:dynamic-gp-average} and \ref{sec:infinite-line-average}
describe the spreading of the quantum density notch through an average
of the normalized densities of GP solitons moving with different velocities.
We now investigate the subsequent temporal fluctuations in the finite
quantum system. We quantify these fluctuations by the temporal variance about the
long-time mean, compare them with predictions from the same GP family, and
examine the time scales determined by the yrast energy differences.

\subsection{Temporal variance of the density profile}
\label{subsec:fluctuation-definition}

We define the temporal variance of the quantum
density profile about its long-time mean,
averaged over the ring.
By expanding the density in spatial Fourier modes,
we express the total variance in terms of
the yrast state energies and density form factors.

Recall the normalized density profile defined in
\eqref{eq:RN-early-def},
\[
 R_N(y,t)=\frac{\rho_N(x_0+\ell_c y,t)}{\bar n},\qquad x_0=\frac L2.
\]
We denote the infinite-time average of a function $f(t)$ by
\[
 \overline f:=\lim_{T\to\infty}\frac1T
 \int_0^T f(t)\,\dd t.
\]
In particular, the infinite-time mean of the density profile is
$\overline R_N(y):=\overline{R_N(y,\cdot)}$.

We define the spatially averaged temporal variance by
\begin{equation}\label{eq:fluctuation-realspace-definition}
 V_{\rm tot,N}(\infty):=
 \lim_{T\to\infty}\frac1{\Lambda T}\int_0^T
 \int_{-\Lambda/2}^{\Lambda/2}
 |R_N(y,t)-\overline R_N(y)|^2\,\dd y\,\dd t.
\end{equation}
We now evaluate the variance using the form-factor expansion
\eqref{eq:density-double-sum}.

For each $m=1,\ldots,N-1$, the $N-m$ pairs of yrast states
$(p,p+m)$, with $p=0,\ldots,N-1-m$, have the same momentum
difference $P_{p+m}-P_p=2\pi m/L$ and hence contribute to
the same spatial Fourier mode.
Recalling the Fourier expansion \eqref{eq:dyn-quantum-full}
and the coefficient formula \eqref{eq:dyn-quantum-B},
we write the normalized density as
\begin{equation}\label{eq:R-Fourier-dynamic-main}
 R_N(y,t)=1+2\mathrm{Re}\sum_{m=1}^{N-1}
 c_{m,N}(t)e^{-2\pi i my/\Lambda},
\end{equation}
where
\begin{equation}\label{eq:cm-time-main}
 c_{m,N}(t):=\frac{(-1)^m}{N\bar n}
 \sum_{p=0}^{N-1-m}F_{p,p+m}
 e^{i(E_{p+m}-E_p)t}.
\end{equation}

Substituting \eqref{eq:R-Fourier-dynamic-main} into
\eqref{eq:fluctuation-realspace-definition}
and using the Fourier orthogonality relation
\begin{equation}\label{eq:spatial-Fourier-orthogonality}
 \frac1\Lambda\int_{-\Lambda/2}^{\Lambda/2}
 e^{2\pi i(m-n)y/\Lambda}\,\dd y
 =\delta_{mn},\qquad m,n\in\mathbb Z,
\end{equation}
we obtain
\begin{equation}\label{eq:fluctuation-mode-sum}
 V_{\rm tot,N}(\infty)=2\sum_{m=1}^{N-1}V_{m,N}(\infty),\qquad
 V_{m,N}(\infty):=\overline{|c_{m,N}|^2}-|\overline c_{m,N}|^2.
\end{equation}
Here $V_{m,N}(\infty)$ is the temporal variance of the
$m$th Fourier coefficient. The factor $2$ accounts for
the equal contributions from the positive and negative
Fourier modes.

To evaluate the individual mode variances, we introduce the
amplitudes $A_{p,m,N}$ and recall the transition frequencies
$\omega_{p,m,N}$ from \eqref{eq:dyn-transition-frequency}:
\begin{equation*}
 A_{p,m,N}:=\frac{(-1)^m}{N\bar n}F_{p,p+m},\qquad
 \omega_{p,m,N}:=E_{p+m}-E_p.
\end{equation*}
Then \eqref{eq:cm-time-main} becomes
\begin{equation}\label{eq:cm-oscillatory-sum}
 c_{m,N}(t)=\sum_{p=0}^{N-1-m}
 A_{p,m,N}e^{i\omega_{p,m,N}t}.
\end{equation}
Taking the squared magnitude, we obtain
\begin{equation}\label{eq:mode-power-expansion}
 |c_{m,N}(t)|^2=
 \sum_{p,q=0}^{N-1-m}A_{p,m,N}A_{q,m,N}^{*}
 e^{i(\omega_{p,m,N}-\omega_{q,m,N})t}.
\end{equation}

If, for each $m$, the frequencies $\omega_{p,m,N}$ are
nonzero and pairwise distinct, taking the infinite-time
average of \eqref{eq:cm-oscillatory-sum} and
\eqref{eq:mode-power-expansion} gives
$\overline c_{m,N}=0$ and
\begin{equation}\label{eq:Vinfty-nondegenerate-main}
 V_{m,N}(\infty)=\sum_{p=0}^{N-1-m}|A_{p,m,N}|^2
 =\frac1{N^2\bar n^2}\sum_{p=0}^{N-1-m}|F_{p,p+m}|^2.
\end{equation}
In this case, $\overline R_N(y)=1$ also follows from
\eqref{eq:R-Fourier-dynamic-main}.
To check the applicability of these results to the systems
studied below, we numerically tested the nonzero and
pairwise distinct frequency conditions for every positive
mode in four parameter sets:
(i) $N=L\in\{40,60,80,100,120,200,300,400,600\}$
at $c=0.01$;
(ii) $N=L\in\{40,60,80,100,120,200,300,400,600\}$
with $c\in\{0.025,0.05,0.1,0.2\}$;
(iii) $N=L\in\{120,200,300,400,600\}$
with $c=400/N^2$ (fixed $\Lambda=20$);
and (iv) $N=L\in\{200,400,600\}$
with $c=16/N^2$ (fixed $\Lambda=4$).

In these numerical checks, frequencies are treated as equal
if their separation is at most
\[
 10^{-12}+10^{-10}\max\bigl(1,\max_p|\omega_{p,m,N}|\bigr),
\]
and as zero if their magnitude is at most this threshold.
No repeated or zero frequencies were detected at this tolerance.
Thus, for these systems, we use
\eqref{eq:Vinfty-nondegenerate-main}, with
$\overline R_N(y)=1$.
Changing the tolerance by factors $0.01$, $10$, and $100$
changes the computed total variances by at most approximately
$0.011\%$, while the frequency classifications at fixed
$\Lambda=20$ remain unchanged.

\subsection{GP predictions for the temporal density variance}
\label{subsec:fluctuation-gp}

We now investigate whether the GP soliton family used to
describe the quantum density profile in
Sections~\ref{sec:quantum-to-gp-average}
and \ref{sec:dynamic-gp-average}
can also provide a semiclassical prediction for its
temporal variance. We examine the Fourier
coefficients of the quantum density profile and the GP soliton average profile.

Using \eqref{eq:R-Fourier-dynamic-main} and
\eqref{eq:cm-oscillatory-sum}, and writing
$q_m=2\pi m/\Lambda$, the quantum density profile takes the form
\begin{equation}\label{eq:variance-quantum-expansion}
 R_N(y,t)=1+2\operatorname{Re}\sum_{m=1}^{N-1}
 \left[\sum_{p=0}^{N-1-m}
 A_{p,m,N}e^{i\omega_{p,m,N}t}\right]e^{-iq_my}.
\end{equation}
Taking the complex conjugate of the Fourier coefficients
in \eqref{eq:dyn-classical-fourier}, we write the
GP soliton average in the same Fourier convention as the quantum density:
\begin{equation}\label{eq:variance-gp-expansion}
 R_{\mathrm{dyn},\Lambda}^{\mathrm{GP}}(y,\tau)
 =1+2\operatorname{Re}\sum_{m=1}^{\infty}
 \left[\frac1\Lambda\int_0^1
 \mathfrak b_{m,\Lambda}^{\rm GP}(s)
 e^{iq_mu(s)\tau}\,\dd s\right]e^{-iq_my},
\end{equation}
where $\tau=c\bar n t$.

For each spatial mode $m$, the quantum Fourier coefficient
in \eqref{eq:variance-quantum-expansion} is a sum over
$N-m$ yrast-state pairs, whereas the corresponding
GP coefficient in \eqref{eq:variance-gp-expansion}
is an integral over the canonical momentum $s$.
This motivates the following heuristic correspondence
based on a Riemann-sum approximation.

We associate each yrast-state pair $(p,p+m)$ with the
midpoint of its normalized momenta,
\[
 s_{p,m,N}:=\frac{p+m/2}{N}.
\]
Since successive midpoints are separated by $1/N$,
a Riemann-sum approximation suggests
\begin{align}\label{eq:variance-gp-riemann}
 &\frac1\Lambda\int_0^1
 \mathfrak b_{m,\Lambda}^{\rm GP}(s)
 e^{iq_mu(s)\tau}\,\dd s
 \simeq\frac1{N\Lambda}\sum_{p=0}^{N-1-m}
 \mathfrak b_{m,\Lambda}^{\rm GP}(s_{p,m,N})
 e^{iq_mu(s_{p,m,N})\tau}.
\end{align}
The sampling points do not cover the entire interval
$0\le s\le1$. The approximation therefore requires
$m/N\to0$, together with control of the discretization
and endpoint errors. Extending the correspondence to
the total variance also requires control of the
contributions from higher modes.

Comparing the Fourier coefficients in
\eqref{eq:variance-quantum-expansion} with the
Riemann-sum approximation \eqref{eq:variance-gp-riemann},
we propose the semiclassical amplitude correspondence
\begin{equation}\label{eq:fluctuation-amplitude-correspondence}
 A_{p,m,N}\simeq\frac1{N\Lambda}
 \mathfrak b_{m,\Lambda}^{\rm GP}(s_{p,m,N}).
\end{equation}

Applying the amplitude correspondence
\eqref{eq:fluctuation-amplitude-correspondence}
to the variance formula \eqref{eq:Vinfty-nondegenerate-main},
and approximating the resulting sum by a Riemann integral
at fixed $m$ and $\Lambda$, we obtain
\begin{align*}
 V_{m,N}(\infty)
 &\simeq\frac1{N^2\Lambda^2}\sum_{p=0}^{N-1-m}
 |\mathfrak b_{m,\Lambda}^{\rm GP}(s_{p,m,N})|^2
 \simeq\frac1{N\Lambda^2}\int_0^1
 |\mathfrak b_{m,\Lambda}^{\rm GP}(s)|^2\,\dd s.
\end{align*}
These approximations require control of the
amplitude errors, discretization errors, and endpoint
contributions.

We define the $N$-independent classical quantity
\begin{equation}\label{eq:A-second-moment}
 G_{m,\Lambda}^{\rm GP}:=\frac2{\Lambda^2}
 \int_0^1|\mathfrak b_{m,\Lambda}^{\rm GP}(s)|^2\,\dd s.
\end{equation}

The resulting finite-ring prediction is
\begin{equation}\label{eq:fluctuation-finite-ring-prediction}
 V_{m,N}(\infty)\simeq\frac{G_{m,\Lambda}^{\rm GP}}{2N}.
\end{equation}

Using \eqref{eq:fluctuation-mode-sum}, we obtain the
corresponding prediction for the total variance,
\begin{equation*}
 V_{\rm tot,N}(\infty)\simeq\frac1N
 \sum_{m\ge1}G_{m,\Lambda}^{\rm GP}.
\end{equation*}
This additionally requires control of the accumulated
mode errors and the classical contributions from
$m\ge N$. We test both the individual-mode and
total-variance predictions below.

We next consider the large-ring limit $\Lambda\to\infty$
of the GP prediction. As discussed in
Section~\ref{sec:infinite-line-average},
the periodic GP Fourier coefficients satisfy
\[
 \mathfrak b_{m_\Lambda,\Lambda}^{\rm GP}(s)
 \longrightarrow B_{\xi(s)}(r),
 \qquad \frac{m_\Lambda}{\Lambda}\longrightarrow r>0,
\]
for almost every $s\in(0,1)$, where $B_{\xi(s)}(r)$
is defined in \eqref{eq:Bxi-main}.
We define
\begin{equation}\label{eq:Phi-def-main}
 \Phi(r):=2\int_0^1|B_{\xi(s)}(r)|^2\,\dd s.
\end{equation}
Using \eqref{eq:Bxi-main} and the measure
\eqref{eq:measure-change-main}, followed by the
substitution $\xi=\cos\alpha$, we obtain
\begin{equation}\label{eq:Phi-integral-main}
 \Phi(r)=16\pi^3r^2\int_0^\pi\sin^2\alpha\,
 \csch^2\!\left(\frac{\pi^2r}{\sin\alpha}\right)
 \dd\alpha.
\end{equation}
With $\Phi(0)$ defined by continuity, we have
\begin{equation}\label{eq:Phi-constants-main}
 \Phi(0)=6,\qquad
 I_1:=\int_0^\infty\Phi(r)\,\dd r
 =\frac{128}{45\pi}.
\end{equation}
These evaluations are given in
Appendix~\ref{app:fluctuation-second-moment}.

By \eqref{eq:A-second-moment} and the convergence
\eqref{eq:app-local-moment-conv}, we have
\[
 \Lambda^2 G_{m_\Lambda,\Lambda}^{\rm GP}
 \longrightarrow \Phi(r).
\]
Thus, the finite-ring prediction
\eqref{eq:fluctuation-finite-ring-prediction}
leads to
\begin{equation}\label{eq:quantum-second-conditional-main}
 V_{m,N}(\infty)\simeq
 \frac{\Phi(m/\Lambda)}{2N\Lambda^2}.
\end{equation}
Combining \eqref{eq:quantum-second-conditional-main}
with \eqref{eq:fluctuation-mode-sum} and approximating
the sum over spatial modes by a Riemann integral
as $\Lambda\to\infty$, we obtain
\begin{align}\label{eq:total-variance-main}
 V_{\rm tot,N}(\infty)
 &\simeq\frac1{N\Lambda^2}
 \sum_{m\ge1}\Phi(m/\Lambda)
 \simeq\frac1{N\Lambda}
 \int_0^\infty\Phi(r)\,\dd r
 =\frac{128}{45\pi N\Lambda}.
\end{align}

In the large-ring GP prediction, approximately $90\%$
of the total variance comes from modes
$m\lesssim0.216\,\Lambda$, since
\[
 \frac1{I_1}\int_0^{0.216}\Phi(r)\,\dd r\simeq0.9.
\]
See Appendix~\ref{app:fluctuation-second-moment}
for the corresponding cumulative mode-sum analysis.
Thus, the relevant modes have $m=O(\Lambda)$.
At fixed $\Lambda$, these modes satisfy
\[
 \frac mN=O\!\left(\frac{\Lambda}{N}\right)
 \longrightarrow0
\]
as $N\to\infty$, consistent with the necessary condition
for the Riemann-sum approximation
\eqref{eq:variance-gp-riemann}.
This motivates testing the variance predictions
along a fixed-$\Lambda$ sequence.

By contrast, at unit density and fixed $c$,
we have $\Lambda=N\sqrt c$, so for fixed
$r=m/\Lambda>0$,
\[
 \frac mN=r\sqrt c.
\]
Increasing $N$ alone therefore does not make
$m/N$ vanish; taking $c\to0$ is also necessary.

\begin{figure}[htbp]
\centering
\includegraphics[width=.78\linewidth]{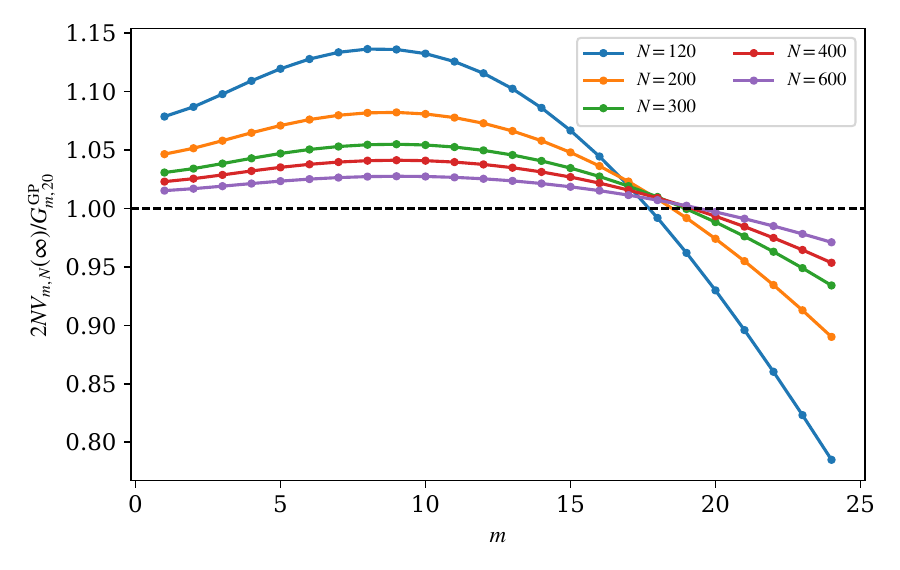}
\caption{Ratios $2NV_{m,N}(\infty)/G_{m,20}^{\rm GP}$ for individual
modes $m\le24$ at $\Lambda=20$, with $N=L=120,200,300,400,600$
and $c=400/N^2$. The horizontal line at one indicates agreement
with the periodic GP prediction.}
\label{fig:finite-ring-variance-main}
\end{figure}

We first test the finite-ring mode prediction at $\Lambda=20$
and unit density, using $N=L=120,200,300,400,600$ and $c=400/N^2$.
Figure~\ref{fig:finite-ring-variance-main} shows the ratios
$2NV_{m,N}(\infty)/G_{m,20}^{\rm GP}$ for individual modes.
At $N=600$, the deviations for $m=1,2,4,8,16$ are approximately
$1.5$--$2.7\%$.

\begin{figure}[htbp]
\centering
\includegraphics[width=\linewidth]{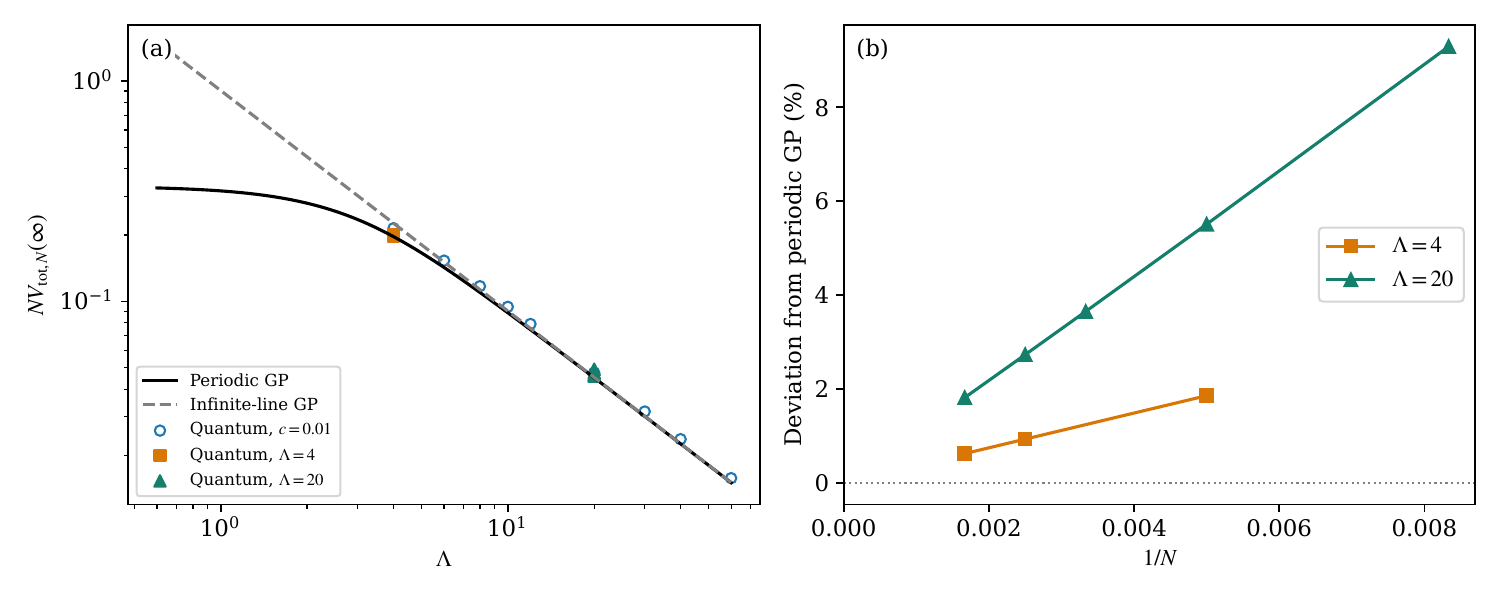}
\caption{Comparison of the total temporal variance
in the Lieb--Liniger model with GP predictions at unit density.
(a) $NV_{\rm tot,N}(\infty)$ versus $\Lambda$,
together with the finite-ring GP result
$S_{\rm GP}(\Lambda)$ and its large-ring
approximation $I_1/\Lambda$.
The values of $NV_{\rm tot,N}(\infty)$ are calculated
at fixed $c=0.01$ and along sequences with
fixed $\Lambda=4$ and $\Lambda=20$.
(b) Relative deviations (in percent),
$100[NV_{\rm tot,N}(\infty)/S_{\rm GP}(\Lambda)-1]$
for $\Lambda=4$ ($N=200,400,600$) and
$\Lambda=20$ ($N=120,200,300,400,600$).
Lines connect the data points.}
\label{fig:ring-variance-comparison}
\end{figure}

For the total variance, we define
\begin{equation}\label{eq:GP-total-variance}
 S_{\rm GP}(\Lambda)
 :=\sum_{m\ge1}G_{m,\Lambda}^{\rm GP}.
\end{equation}
The finite-ring GP prediction is
\[
 NV_{\rm tot,N}(\infty)\simeq S_{\rm GP}(\Lambda),
\]
and its large-ring limit satisfies
\[
 \Lambda S_{\rm GP}(\Lambda)\longrightarrow I_1=\frac{128}{45\pi}
 \qquad (\Lambda\to\infty).
\]
Figure~\ref{fig:ring-variance-comparison}
compares the rescaled quantum total variance
$NV_{\rm tot,N}(\infty)$ with the finite-ring
GP prediction $S_{\rm GP}(\Lambda)$
and the large-ring approximation $I_1/\Lambda$.
The relative difference between $S_{\rm GP}(\Lambda)$
and $I_1/\Lambda$, measured with respect to
$I_1/\Lambda$, is about $13\%$ at $\Lambda=4$
and only $0.41\%$ at $\Lambda=20$.

We evaluate the quantum total variance along two
fixed-$\Lambda$ sequences:
$N=L=120,200,300,400,600$ with $c=400/N^2$
at $\Lambda=20$, and
$N=L=200,400,600$ with $c=16/N^2$
at $\Lambda=4$.
At $\Lambda=20$, the relative deviation of
$NV_{\rm tot,N}(\infty)$ from $S_{\rm GP}(20)$
decreases from about $9.3\%$ at $N=120$
to $1.8\%$ at $N=600$.
At $\Lambda=4$, the corresponding deviations
are about $1.9\%$, $0.93\%$, and $0.62\%$.
Both sequences show improving agreement with
$S_{\rm GP}(\Lambda)$ as $N$ increases.
At $\Lambda=4$ and $N=600$, the first Fourier mode
accounts for about $93.5\%$ of the total variance,
consistent with the corresponding GP result.
The ultraweak-coupling condition and the small-$\Lambda$
check are discussed in
Appendix~\ref{app:fluctuation-second-moment}.

\begin{figure}[htbp]
\centering
\includegraphics[width=.78\linewidth]{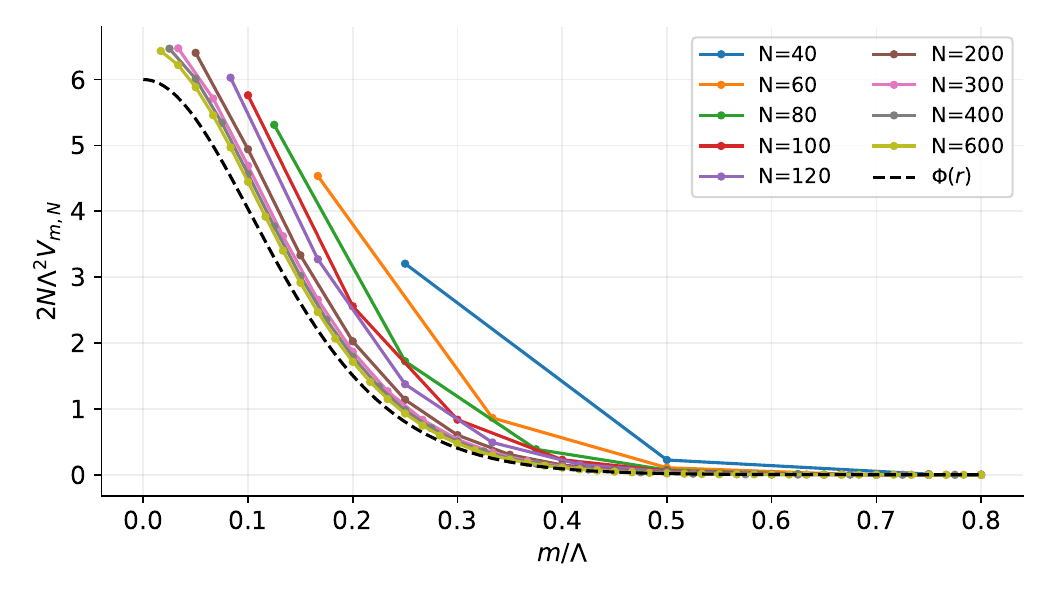}
\caption{Comparison of the Lieb--Liniger mode variances
with the large-ring GP prediction at $c=0.01$,
$\bar n=1$, and $N=L=40,60,80,100,120,200,300,400,600$.
The vertical axis is $2N\Lambda^2V_{m,N}(\infty)$,
and the dashed curve is $\Phi(m/\Lambda)$.}
\label{fig:mode-variance-main}
\end{figure}

We now test the large-ring mode-variance prediction
\eqref{eq:quantum-second-conditional-main}
at fixed $c=0.01$ and unit density, using
$N=L=40,60,80,100,120,200,300,400,600$.
Here $\Lambda=0.1N$ ranges from $4$ to $60$,
so the smallest systems are not in the large-ring regime.
Figure~\ref{fig:mode-variance-main} compares
$2N\Lambda^2V_{m,N}(\infty)$ with $\Phi(m/\Lambda)$.
At $N=600$, the cumulative variance in the
Lieb--Liniger model reaches $90\%$ at $m=13$,
in close agreement with the GP prediction
$m\simeq0.216\Lambda\simeq13$ from the cumulative
mode distribution \eqref{eq:app-cumulative-GP}.

We next examine the total variance along the same
fixed-$c$ sequence. At unit density, $\Lambda=N\sqrt c$,
so \eqref{eq:total-variance-main} becomes
\[
 V_{\rm tot,N}(\infty)\simeq\frac{I_1}{N^2\sqrt c},
 \qquad I_1=\frac{128}{45\pi}.
\]

\begin{figure}[htbp]
\centering
\includegraphics[width=.94\linewidth]{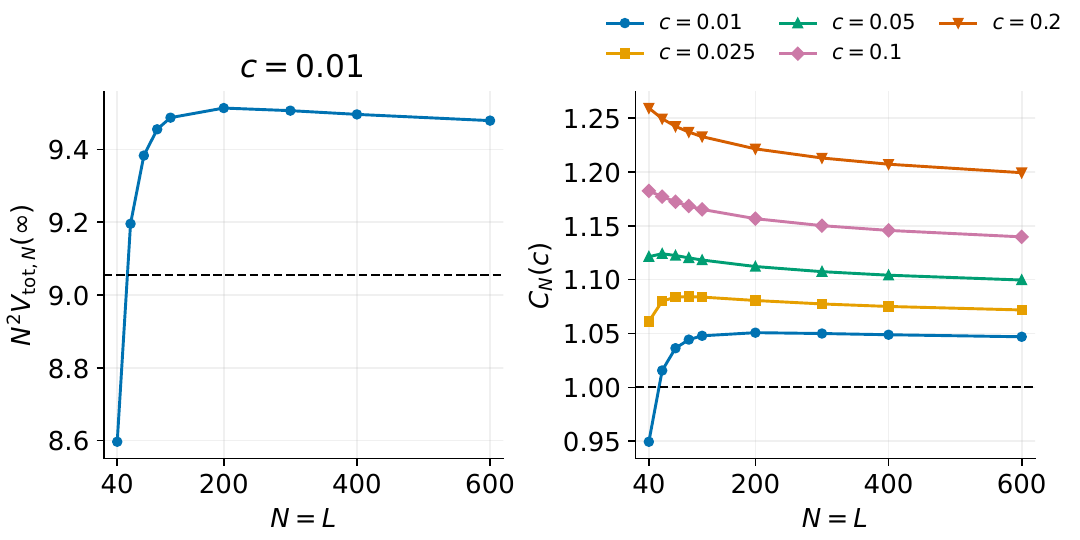}
\caption{Comparison of the total temporal variance in the
Lieb--Liniger model with the large-ring GP prediction
at unit density.
Left: $N^2V_{\rm tot,N}(\infty)$ for $c=0.01$,
compared with the GP prediction $I_1/\sqrt c$ (dashed).
Right: the ratio $C_N(c)$ for
$c=0.01,0.025,0.05,0.1,0.2$.
All series use $N=L=40,60,80,100,120,200,300,400,600$
and include all modes $m=1,\ldots,N-1$.
The horizontal dashed line in the right panel
indicates agreement with the GP prediction.}
\label{fig:total-variance-comp-main}
\end{figure}

We test both the predicted $N^{-2}$ dependence and its coefficient.
The left panel of Fig.~\ref{fig:total-variance-comp-main} plots
$N^2V_{\rm tot,N}$, which should approach the constant $I_1/\sqrt c$
if the prediction holds.
At $c=0.01$, the values obtained from the Lieb--Liniger
model change only from approximately $9.51$ at $N=200$
to $9.48$ at $N=600$, consistent with an approximately
$N^{-2}$ dependence over this range.
The GP prediction for the constant, however, is
approximately $9.05$.

To examine the agreement with the predicted coefficient,
we introduce the ratio
\begin{equation*}
 C_N(c):=\frac{V_{\rm tot,N}(\infty)}{I_1/(N^2\sqrt c)}
 =\frac{N\Lambda V_{\rm tot,N}(\infty)}{I_1}.
\end{equation*}
The right panel of Figure~\ref{fig:total-variance-comp-main}
plots $C_N(c)$, with $C_N(c)=1$ corresponding to agreement
with the GP prediction.
At $c=0.01$ and $N=600$, the variance in the
Lieb--Liniger model remains approximately $4.7\%$
above the GP prediction.

For the other couplings $c=0.025,0.05,0.1,0.2$,
the ratios decrease with $N$ over the sizes studied,
but at $N=600$ the variances still exceed the GP
predictions by approximately $7.2\%$, $10\%$,
$14\%$, and $20\%$, respectively.
The agreement is better at weaker coupling.
At fixed $c$, however, increasing $N$ leaves the
dimensionless coupling $\gamma=c$ unchanged,
so this limit does not approach the weak-coupling regime.

\subsection{Transition frequencies and initial dephasing}
\label{subsec:fluctuation-dephasing}

We now examine the initial dephasing of the Fourier coefficients
$c_{m,N}(t)$ in \eqref{eq:cm-oscillatory-sum}. Contributions from different
yrast-state pairs generally have different transition frequencies and
can lose phase alignment.
We first relate the transition frequencies of the Lieb-Liniger model
to the GP soliton
velocities and then test whether the initial decay of
$|c_{m,N}(t)|^2$ occurs on a time scale set by the inverse
of the weighted frequency width.

To relate the transition frequencies to GP motion,
we introduce a smooth interpolation $E(P)$ of the
discrete yrast spectrum, with $E(P_p)=E_p$,
and write $v(P)=\dd E/\dd P$.
When the momentum difference
$P_{p+m}-P_p=2\pi m/L$ is sufficiently small
on the scale over which the dispersion varies,
\begin{equation}\label{eq:fluctuation-dispersion-frequency}
 \frac{E_{p+m}-E_p}{P_{p+m}-P_p}
 \simeq v\!\left(\frac{P_p+P_{p+m}}2\right).
\end{equation}

In the weak-coupling semiclassical regime, the
energy--momentum dispersion of the Lieb--Liniger
yrast states is expected to approach that of the
periodic GP solitons \cite{KCU2009,KCU2010}.
For GP solitons, the derivative of the energy
with respect to the canonical momentum equals
the physical propagation velocity,
$dE_{\rm GP}/dP_{\rm GP}=v_{\rm GP}$.
Since the GP profile
$R_{s,\Lambda}^{\rm per}(y-u(s)\tau)$
moves with dimensionless velocity $u(s)$,
the coordinate transformations in
\eqref{eq:prelim-healing} give
\[
 v_{\rm GP}(s)
 =\frac{dx}{dt}
 =\ell_c\frac{dy}{d\tau}\frac{d\tau}{dt}
 =\ell_c c\bar n\,u(s)
 =\sqrt{c\bar n}\,u(s).
\]
Here $u(s)$ is defined in \eqref{eq:GP-velocity-s},
and $s=P_{\rm GP}/(2\pi\bar n)$ by
\eqref{eq:proposal-sdef}.
Assuming the corresponding agreement of the
dispersion slopes, we therefore expect
\[
 v(P)\simeq\sqrt{c\bar n}\,
 u\left(\frac{P}{2\pi\bar n}\right).
\]

Using $P_p=2\pi p/L$ and $\bar n=N/L$,
the midpoint momentum corresponds to
$s_{p,m,N}=(p+m/2)/N$, as introduced in
the preceding subsection.
Combining this with
\eqref{eq:fluctuation-dispersion-frequency}, we obtain
\begin{align}\label{eq:fluctuation-gp-frequency}
 \frac{\omega_{p,m,N}}{c\bar n}
 &\simeq
 \frac{P_{p+m}-P_p}{c\bar n}
 v\!\left(\frac{P_p+P_{p+m}}2\right)
\simeq
 \frac{2\pi m}{L\sqrt{c\bar n}}\,
 u(s_{p,m,N})
 =q_m u(s_{p,m,N}),
\end{align}
where we used $L=\ell_c\Lambda$ and
$\ell_c=(c\bar n)^{-1/2}$.
Thus, the quantum phase factors
$e^{i\omega_{p,m,N}t}$ in
\eqref{eq:variance-quantum-expansion}
correspond semiclassically to the GP phase factors
$e^{iq_mu(s_{p,m,N})\tau}$,
where $\tau=c\bar n t$.

We next characterize the spread of the transition
frequencies $\omega_{p,m,N}=E_{p+m}-E_p$
within each spatial mode $m$.
To avoid having frequencies with negligible contributions
determine the width, we use the squared amplitudes
$|A_{p,m,N}|^2$ as weights.
Let $p_j$ denote the state-pair indices arranged
in increasing order of transition frequency:
\begin{equation}\label{eq:ordered-transition-frequencies-main}
 \Omega_{j,m,N}:=\omega_{p_j,m,N},\qquad
 \Omega_{1,m,N}<\cdots<\Omega_{N-m,m,N}.
\end{equation}
For fixed $m,N$, we abbreviate $\Omega_{j,m,N}$
by $\Omega_j$ and define
\begin{equation*}
 w_j:=|A_{p_j,m,N}|^2,\qquad
 \pi_j:=\frac{w_j}{\sum_{a=1}^{N-m}w_a}.
\end{equation*}
By \eqref{eq:Vinfty-nondegenerate-main}, $\pi_j$
is the fraction of the mode variance $V_{m,N}(\infty)$
contributed
by the pair $(p_j,p_j+m)$.

For the weighted percentiles, we use
\begin{equation*}
 h_j:=\sum_{a<j}\pi_a+\frac{\pi_j}{2}.
\end{equation*}
We then interpolate linearly between
$(h_j,\Omega_j)$ and $(h_{j+1},\Omega_{j+1})$:
\[
 \Omega_{m,N}^{[q]}
 =\Omega_j+\frac{q-h_j}{h_{j+1}-h_j}
 (\Omega_{j+1}-\Omega_j),
 \qquad h_j\le q\le h_{j+1}.
\]
We use constant extensions outside the interval
$[h_1,h_{N-m}]$. If there is only one contributing
term, all percentiles equal its frequency.
We define the weighted frequency width by
\begin{equation}\label{eq:bandwidth-main}
 \Delta\Omega_{m,N}:=
 \Omega_{m,N}^{[0.90]}-\Omega_{m,N}^{[0.10]}.
\end{equation}

\begin{figure}[htbp]
\centering
\includegraphics[width=.88\linewidth]{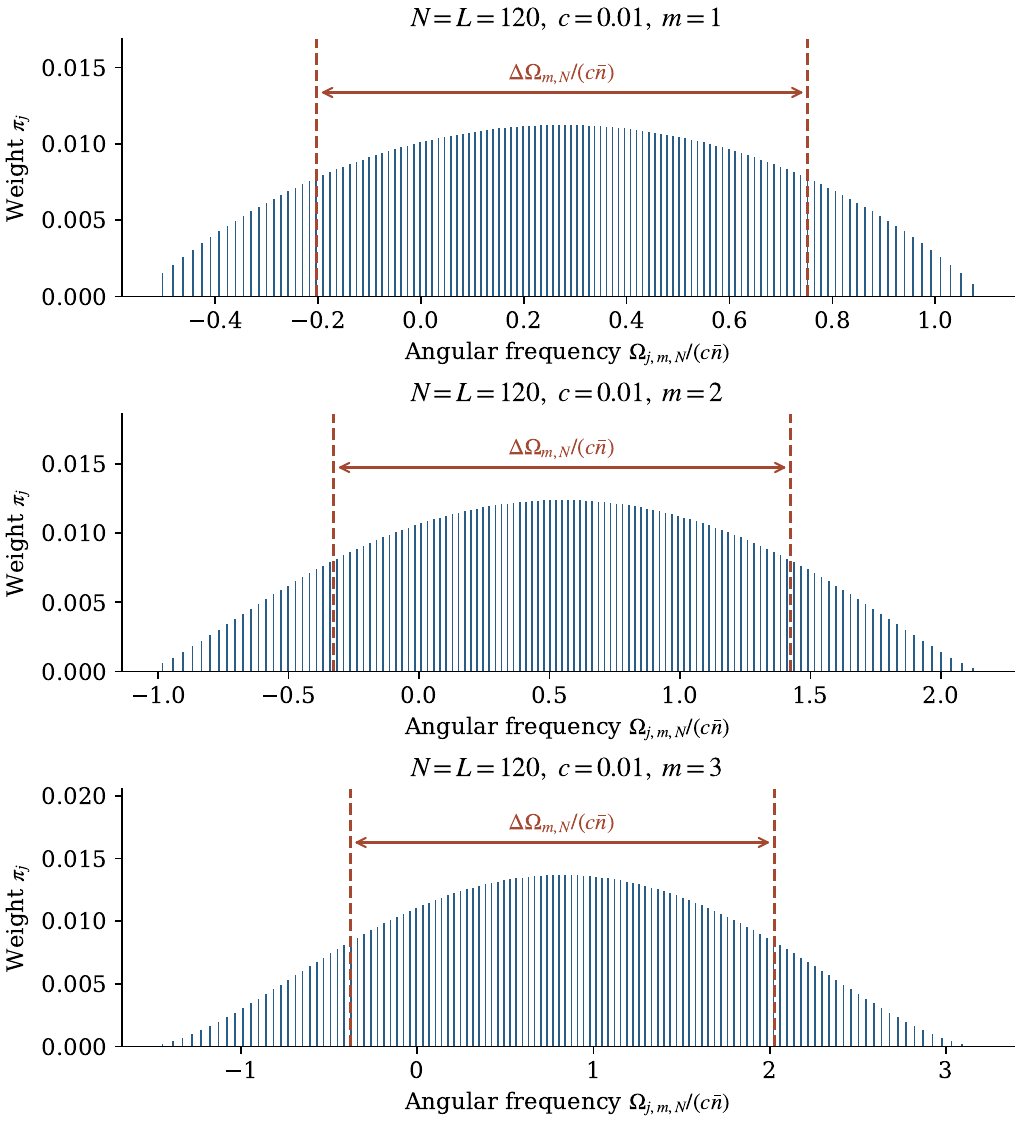}
\caption{Weighted angular-frequency distributions
in the Lieb--Liniger model at $N=L=120$ and $c=0.01$,
for the three lowest positive modes $m=1,2,3$.
Line heights are the normalized squared-amplitude
weights $\pi_j$.
Dashed lines mark $\Omega^{[0.10]}_{m,N}/(c\bar n)$
and $\Omega^{[0.90]}_{m,N}/(c\bar n)$;
the arrow indicates their separation
$\Delta\Omega_{m,N}/(c\bar n)$.}
\label{fig:actual-frequency-distributions}
\end{figure}

Figure~\ref{fig:actual-frequency-distributions} displays
the weighted distributions of the transition frequencies
for
$N=L=120$ and $c=0.01$.
We show the three lowest positive modes $m=1,2,3$,
which cover the main low-mode region of the GP variance
spectrum discussed in the preceding subsection.
Each vertical line represents one yrast-state pair
and has height $\pi_j$.
The dashed lines mark the weighted 10th and 90th
percentiles, whose separation gives
$\Delta\Omega_{m,N}$.

\FloatBarrier

The weighted frequency width $\Delta\Omega_{m,N}$
suggests a characteristic initial dephasing time of
order $1/\Delta\Omega_{m,N}$~\cite{deOliveira2018}.
We test this prediction using the time-dependent
Fourier coefficients in \eqref{eq:cm-oscillatory-sum}.
By \eqref{eq:Vinfty-nondegenerate-main},
the long-time mean of $|c_{m,N}(t)|^2$
is $V_{m,N}(\infty)$.
When the initial squared amplitude differs from
this mean, we define
\begin{equation}\label{eq:mode-collapse-main}
 \mathcal D_{m,N}(t):=
 \frac{|c_{m,N}(t)|^2-V_{m,N}(\infty)}
 {|c_{m,N}(0)|^2-V_{m,N}(\infty)}.
\end{equation}
This normalization gives $\mathcal D_{m,N}(0)=1$
and $\overline{\mathcal D_{m,N}}=0$.

To compare dephasing across different system sizes
at approximately the same spatial wave numbers,
we select the positive integer modes $m$ for which
$m/\Lambda$ is closest to $0.08$, $0.16$, and $0.24$.
These values sample the low-mode region in which
the GP variance spectrum is concentrated, as
discussed in the preceding subsection.
For small $\Lambda$, different target values may
select the same integer mode, which is then counted
only once for each system.
For $c=0.01$, we use the 21 pairs $(m,N)$
listed in
Table~\ref{tab:dephasing-selected-modes}.

\begin{table}[htbp]
\centering
\begin{tabular}{ccc}
\hline
$N=L$ & $\Lambda$ & Selected modes $m$ \\
\hline
40  & 4  & $1$ \\
60  & 6  & $1$ \\
80  & 8  & $1,2$ \\
100 & 10 & $1,2$ \\
120 & 12 & $1,2,3$ \\
200 & 20 & $2,3,5$ \\
300 & 30 & $2,5,7$ \\
400 & 40 & $3,6,10$ \\
600 & 60 & $5,10,14$ \\
\hline
\end{tabular}
\caption{Fourier modes selected for the dephasing
comparison at $c=0.01$ and unit density,
with $\Lambda=0.1N$.}
\label{tab:dephasing-selected-modes}
\end{table}

\begin{figure}[htbp]
\centering
\includegraphics[width=.94\linewidth]{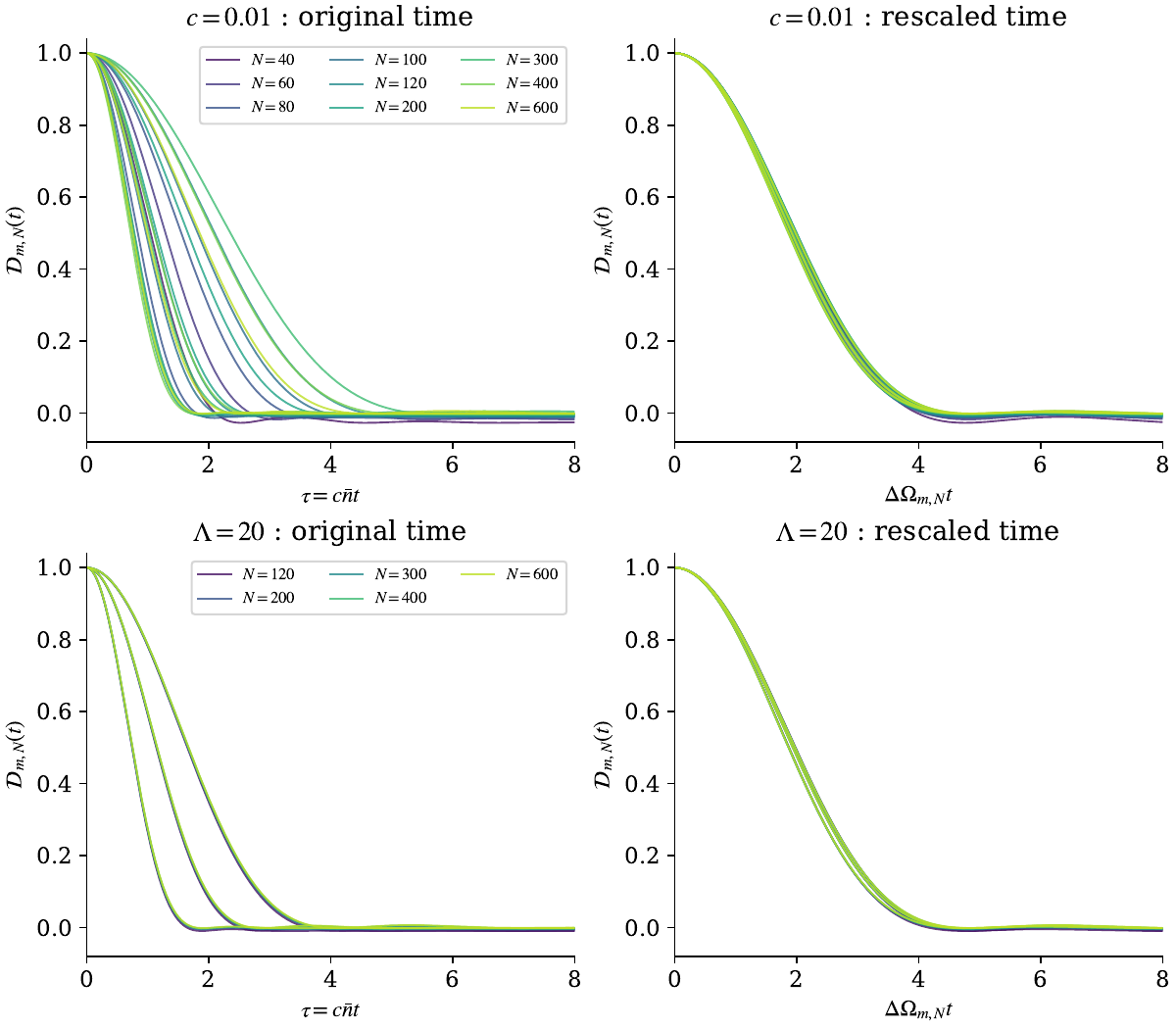}
\caption{Initial decay of the normalized squared
Fourier amplitude $\mathcal D_{m,N}(t)$
\eqref{eq:mode-collapse-main}.
Top: fixed $c=0.01$, with the 21 pairs $(N,m)$
listed in Table~\ref{tab:dephasing-selected-modes}.
Bottom: fixed $\Lambda=20$, with
$N=L=120,200,300,400,600$ and $m=2,3,5$
for each system.
Each curve corresponds to one pair $(N,m)$,
with colors indicating $N$.
Left: scaled time $\tau=c\bar n t$.
Right: rescaled time $\Delta\Omega_{m,N}t$.
}
\label{fig:bethe-mode-collapse-main}
\end{figure}

Figure~\ref{fig:bethe-mode-collapse-main} 
compares
the decay curves $\mathcal{D}_{m,N}(t)$
before and after rescaling time.
The top row shows the 21 curves corresponding
to the pairs $(m,N)$ listed in
Table~\ref{tab:dephasing-selected-modes}.
The bottom row shows the fixed-$\Lambda=20$ sequence
with $N=L=120,200,300,400,600$, for which
the selected modes are $m=2,3,5$ at each size,
giving 15 curves.
Each curve corresponds to a fixed pair $(m,N)$.
The left panels use the scaled time $\tau=c\bar n t$,
while the right panels use the rescaled time
$\Delta\Omega_{m,N}t$, with the frequency width
$\Delta\Omega_{m,N}$ calculated separately
for each pair $(m,N)$.

After rescaling, the decay curves nearly collapse
onto a common profile.
This suggests that, over the initial-decay regime,
\begin{equation}\label{eq:dephasing-collapse-suggestion}
 \mathcal D_{m,N}(t)
 \approx \mathcal F(\Delta\Omega_{m,N}t),
\end{equation}
where $\mathcal F$ is approximately independent
of $m$ and $N$ for the modes and system sizes
examined here.

To quantify the observed collapse, let $t_{1/2}$
denote the first time at which
$\mathcal D_{m,N}(t)=1/2$.
For the 21 fixed-$c=0.01$ curves, the rescaled
half-decay times $\Delta\Omega_{m,N}t_{1/2}$
have a mean of $1.94$ and a standard deviation
of $0.05$.
This small variation further supports the existence
of an approximately common initial-decay profile,
as suggested by \eqref{eq:dephasing-collapse-suggestion}.

\subsection{Discrete frequency spacing and finite observation times}
\label{subsec:fluctuation-window}

The preceding subsection examined the initial dephasing of individual
Fourier modes. We now ask how the variance measured over a finite
observation interval $[0,T]$ approaches its infinite-time value.

For a function $f(t)$, we define its average
over the interval $[0,T]$ by
\[
 \langle f\rangle_T:=\frac1T\int_0^T f(t)\,\dd t.
\]
The variance of the $m$th Fourier coefficient over $[0,T]$ is
\begin{align}\label{eq:finite-window-var-main}
 V_{m,N}[0,T]
 &:=\frac1T\int_0^T
 |c_{m,N}(t)-\langle c_{m,N}\rangle_T|^2\,\dd t
 =\langle|c_{m,N}|^2\rangle_T
   -|\langle c_{m,N}\rangle_T|^2.
\end{align}

To evaluate this variance, we use the Fourier-coefficient expansion
\eqref{eq:cm-oscillatory-sum} and introduce the finite-time averaging
factor
\begin{equation}\label{eq:WT-main}
 W_T(a):=\langle e^{iat} \rangle_T=\frac1T\int_0^T e^{iat}\,\dd t
 =e^{iaT/2}\frac{\sin(aT/2)}{aT/2},\qquad W_T(0)=1.
\end{equation}
Substituting the Fourier coefficient expansion
\eqref{eq:cm-oscillatory-sum} into
\eqref{eq:finite-window-var-main}
and using \eqref{eq:WT-main}, we obtain
\begin{align}\label{eq:finite-window-kernel-main}
 V_{m,N}[0,T]
 &=\sum_{p,q=0}^{N-1-m}A_{p,m,N}A_{q,m,N}^{*}\nonumber\\
 &\quad\times\bigl[
 W_T(\omega_{p,m,N}-\omega_{q,m,N})
 -W_T(\omega_{p,m,N})W_T(\omega_{q,m,N})^{*}
 \bigr].
\end{align}
The first term in the brackets depends on differences
of transition frequencies, whereas the second term
arises from $|\langle c_{m,N}\rangle_T|^2$
and depends on the individual frequencies.
In particular, for $a\ne0$,
\[
 |W_T(a)|\le\min\!\left\{1,\frac{2}{|a|T}\right\}.
\]
For $p\ne q$, the factor
$W_T(\omega_{p,m,N}-\omega_{q,m,N})$
becomes small when
$|\omega_{p,m,N}-\omega_{q,m,N}|T\gg1$.
Hence, smaller frequency differences require longer
observation times for these contributions to become small.
Similarly, the factors $W_T(\omega_{p,m,N})$
in the second term become small when
$|\omega_{p,m,N}|T\gg1$.
The weighted frequency width $\Delta\Omega_{m,N}$
studied in the preceding subsection characterizes
initial dephasing, but does not capture the
smaller differences between neighboring frequencies
that can remain significant in the first term
over longer observation times.
We therefore examine these neighboring frequency
differences.

Using the ordered frequencies $\Omega_j$ and
the corresponding indices $p_j$ from
\eqref{eq:ordered-transition-frequencies-main},
we define the adjacent frequency gaps and their weights by
\[
 d_j:=\Omega_{j+1}-\Omega_j,\qquad
 a_j:=|A_{p_j,m,N}A_{p_{j+1},m,N}^{*}|.
\]
The weight $a_j$ is the magnitude of the coefficient
multiplying $W_T(\Omega_j-\Omega_{j+1})$
in \eqref{eq:finite-window-kernel-main}.
To reduce the influence of weakly weighted tails,
we retain the adjacent frequency gaps $d_j$ satisfying
\[
 \Omega_{m,N}^{[0.05]}\le\Omega_j<\Omega_{j+1}
 \le\Omega_{m,N}^{[0.95]},
\]
where the weighted percentiles were defined in the
preceding subsection.

Let $M$ denote the number of retained adjacent gaps.
We arrange these gaps in increasing order,
writing $g_a=d_{j_a}$ with
$g_1\le\cdots\le g_M$, and define
\[
 \rho_a:=\frac{a_{j_a}}{\sum_{b=1}^M a_{j_b}},
 \qquad
 H_a:=\sum_{b<a}\rho_b+\frac{\rho_a}{2}.
\]
Following the interpolation procedure used for the
weighted percentiles in the preceding subsection,
we interpolate linearly through $(H_a,g_a)$
and take the value at cumulative weight $1/2$.
Thus, for $H_a\le1/2\le H_{a+1}$, we define
\begin{equation*}
 \delta\Omega_{m,N}:=
 g_a+\frac{1/2-H_a}{H_{a+1}-H_a}(g_{a+1}-g_a).
\end{equation*}
For a single retained gap, we set
$\delta\Omega_{m,N}=g_1$.
This defines a weighted median of the selected
adjacent gaps.

The weighted frequency width $\Delta\Omega_{m,N}$
defined in \eqref{eq:bandwidth-main} characterizes
the initial dephasing studied in the preceding
subsection.
By contrast, the representative adjacent spacing
$\delta\Omega_{m,N}$ characterizes a longer time
scale associated with the suppression of
contributions from neighboring frequencies.
We therefore introduce
\begin{equation*}
 T_{\rm coarse}(m,N):=\frac1{\Delta\Omega_{m,N}},
 \qquad
 T_{\rm resolve}(m,N):=\frac1{\delta\Omega_{m,N}}.
\end{equation*}

The frequency correspondence
\eqref{eq:fluctuation-gp-frequency},
\[
 \omega_{p,m,N}
 \simeq
 \frac{2\pi m\sqrt{c\bar n}}{L}
 u(s_{p,m,N}),
\]
provides a semiclassical estimate of these two time scales.
Assuming that the variation of $u(s)$ over the momenta
carrying appreciable weight is of order one,
the weighted frequency width $\Delta\Omega_{m,N}$
is estimated as
\begin{equation*}
 \Delta\Omega_{m,N}
 =O\!\left(\frac{m\sqrt{c\bar n}}{L}\right).
\end{equation*}
Assuming $u$ is smooth,
the same correspondence
gives
\begin{equation*}
 \begin{aligned}
 \omega_{p+1,m,N}-\omega_{p,m,N}
 &\simeq c\bar n\,q_m
 \left[u\!\left(s_{p,m,N}+\frac1N\right)-u(s_{p,m,N})\right]\\
 &\simeq\frac{2\pi m\sqrt{c\bar n}}{LN}\,u'(s_{p,m,N}).
 \end{aligned}
\end{equation*}
If $|u'(s)|$ is typically of order one in the
weighted range, the representative spacing has scale
\begin{equation} \label{eq:two-frequency-scales-main}
 \delta\Omega_{m,N}=O\!\left(\frac{m\sqrt{c\bar n}}{LN}\right).
\end{equation}

Based on these heuristic estimates,
$T_{\rm resolve}/T_{\rm coarse}$ is expected
to grow proportionally to $N$. In the dimensionless time
$\tau=c\bar n t$, write
$\tau_{\rm coarse}(m,N):=c\bar n T_{\rm coarse}(m,N)$ and
$\tau_{\rm resolve}(m,N):=c\bar n T_{\rm resolve}(m,N)$.
The corresponding heuristic estimates are
\begin{equation*}
 \tau_{\rm coarse}(m,N)=O(\Lambda/m),\qquad
 \tau_{\rm resolve}(m,N)=O(N\Lambda/m).
\end{equation*}
For modes with $m/\Lambda$ bounded above and below
by positive constants, these estimates reduce to
$\tau_{\rm coarse}=O(1)$ and
$\tau_{\rm resolve}=O(N)$.

\begin{figure}[htbp]
\centering
\includegraphics[width=.94\linewidth]{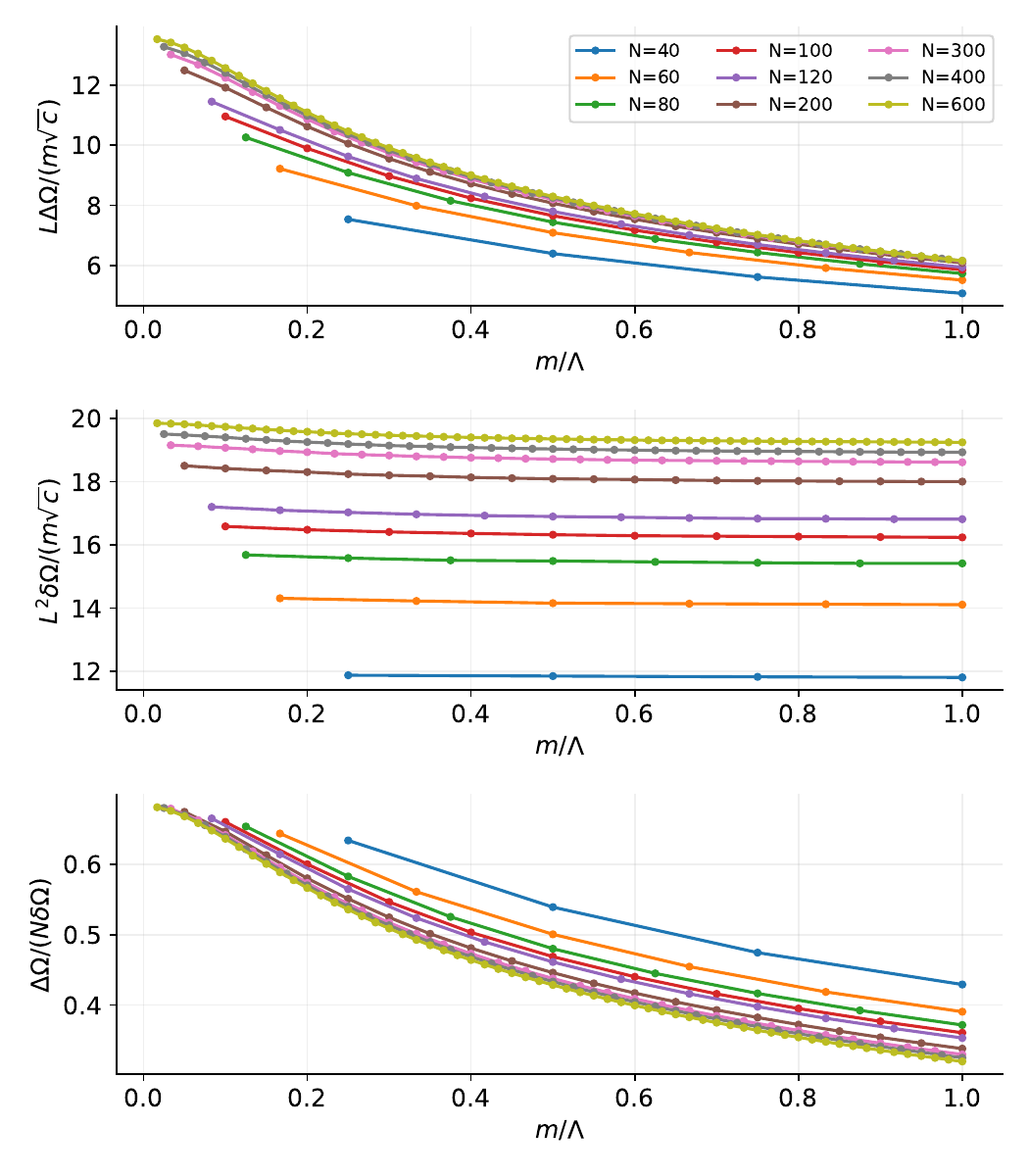}
\caption{Frequency scales at $c=0.01$ for
$N=L=40,60,80,100,120,200,300,400,600$.
From top to bottom: the scaled weighted width
$L\Delta\Omega_{m,N}/(m\sqrt c)$, the scaled representative
adjacent spacing $L^2\delta\Omega_{m,N}/(m\sqrt c)$,
and the normalized time-scale ratio
$\Delta\Omega_{m,N}/(N\delta\Omega_{m,N})$.
The horizontal coordinate is $m/\Lambda$.
Modes satisfy $m/\Lambda\le1$ and have at least five frequencies
in the weighted 5--95 percent interval used to select adjacent gaps.}
\label{fig:time-scale-ratio-main}
\end{figure}

We test the two frequency scales numerically at unit density and
fixed $c=0.01$. Figure~\ref{fig:time-scale-ratio-main} displays the
scaled width $L\Delta\Omega_{m,N}/(m\sqrt c)$, the scaled adjacent
spacing $L^2\delta\Omega_{m,N}/(m\sqrt c)$, and their ratio
$\Delta\Omega_{m,N}/(N\delta\Omega_{m,N})$ as functions of
$r=m/\Lambda$.
At a given $r$, both scaled quantities increase with $N$,
and the larger-system curves lie closer together.
The scaled width decreases with $r$, while the scaled
spacing depends more weakly on $r$.

The bottom panel shows that
$\Delta\Omega_{m,N}/(N\delta\Omega_{m,N})$
lies approximately between $0.32$ and $0.68$
over the plotted range.
For example, at $r=0.5$, this ratio changes
from about $0.54$ at $N=40$ to $0.43$ at $N=600$.
Correspondingly,
$T_{\rm resolve}/T_{\rm coarse}$ increases
from about $22$ to $257$.
These results support the predicted growth
of the time-scale ratio proportionally to $N$.

The half-decay times $t_{1/2}$ from the preceding subsection
provide a further comparison of the two time scales.
For the 21 pairs $(m,N)$ listed in
Table~\ref{tab:dephasing-selected-modes},
the mean of $\delta\Omega_{m,N}t_{1/2}$
over the selected modes at each $N$
decreases from about $0.078$ at $N=40$
to $0.0054$ at $N=600$, whereas
$\Delta\Omega_{m,N}t_{1/2}$ remains close to two.
Thus, when $\mathcal D_{m,N}(t)$ \eqref{eq:mode-collapse-main} first reaches $1/2$,
$\delta\Omega_{m,N}t_{1/2}$ is still much smaller
than one.
This shows that initial dephasing occurs on a
time scale much shorter than $T_{\rm resolve}$.

We finally examine the total density variance measured over a finite
observation interval. By the spatial Fourier orthogonality relation
\eqref{eq:spatial-Fourier-orthogonality} and the definition
\eqref{eq:finite-window-var-main}, we define the finite-window variance by
\begin{align*}
 V_{\rm tot,N}[0,T]
 &:=\frac1\Lambda\int_{-\Lambda/2}^{\Lambda/2}
 \left\langle|R_N(y,t)-\langle R_N(y,\cdot)\rangle_T|^2\right\rangle_T\dd y
 =2\sum_{m=1}^{N-1}V_{m,N}[0,T].
\end{align*}
For comparison with the infinite-time variance,
we consider the ratio
\begin{equation*}
 \frac{V_{\rm tot,N}[0,T]}{V_{\rm tot,N}(\infty)}.
\end{equation*}

At unit density, the estimate \(\eqref{eq:two-frequency-scales-main}\) suggests a characteristic time scale \(T_{\rm resolve}(m,N)\) of order \(L^2/(m\sqrt c)\).
We therefore plot the ratio against
\[
 \zeta:=\frac{T\sqrt c}{L^2},
\]
corresponding to the predicted reference scale for $m=1$.

\begin{figure}[htbp]
\centering
\includegraphics[width=.94\linewidth]{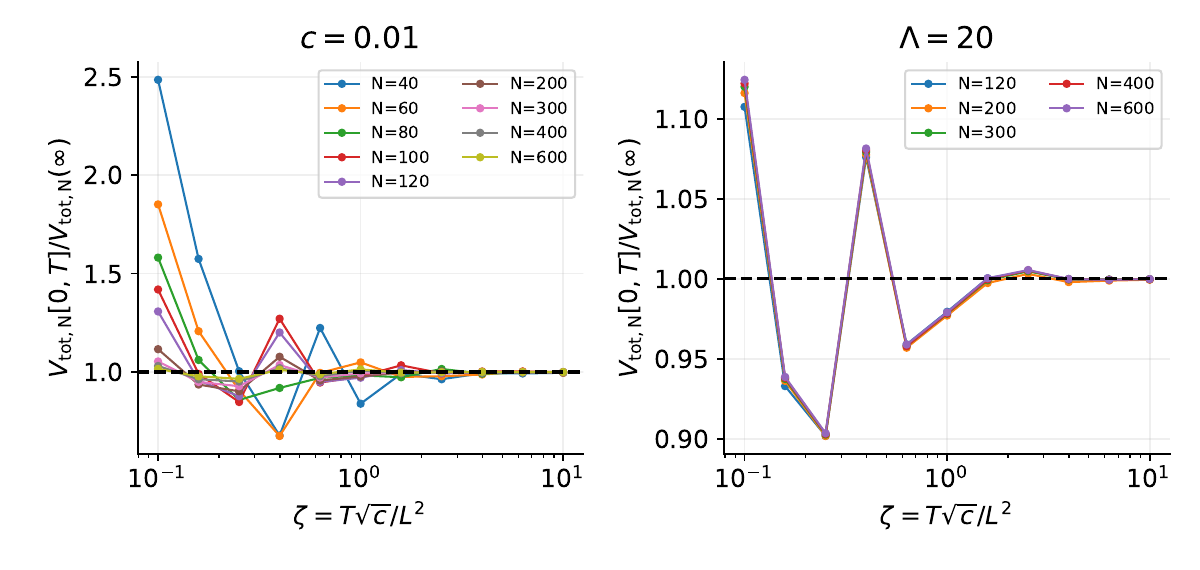}
\caption{
Total finite-window variance normalized
by its infinite-time value.
Left: fixed $c=0.01$ and
$N=L=40,60,80,100,120,200,300,400,600$.
Right: fixed $\Lambda=20$ and $N=L=120,200,300,400,600$.
The horizontal coordinate is $\zeta=T\sqrt c/L^2\ge0.1$.
The numerator is computed from \eqref{eq:finite-window-kernel-main}
and summed over all modes $m=1,\ldots,N-1$, with the mean over
$[0,T]$ subtracted. Dashed lines indicate equality with the
infinite-time variance.}
\label{fig:full-density-ratio-main}
\end{figure}

Figure~\ref{fig:full-density-ratio-main}
shows the ratio of the finite-window variance
$V_{\rm tot,N}[0,T]$ to its infinite-time value
$V_{\rm tot,N}(\infty)$
for fixed
$c=0.01$ and fixed $\Lambda=20$.
For fixed $c=0.01$, smaller systems show larger
deviations from one at relatively short observation times.
For example, at $\zeta=0.1$, the ratio decreases
from approximately $2.5$ at $N=40$ to $1.02$ at $N=600$.
For each $N$, the ratio approaches one as
the observation time increases.
For fixed $\Lambda=20$ with $c=400/N^2$, the five curves nearly
coincide.

\section{Discussion and conclusions}\label{sec:conclusion}

We have investigated the semiclassical description of the equal-weight
quantum dark-soliton states and their time evolution in the Lieb--Liniger
model. Our results support the canonical-momentum average of normalized
periodic GP densities as the semiclassical description of the initial
quantum density. Averaging these density profiles as they move with different soliton
velocities also describes the spreading of the quantum density notch. We also obtained the
infinite-line averaged initial profile and its Fourier transform, including
the algebraic decay of the initial density deficit. See also \cite{SarviVianelloCaux2026}
for recent related work on quantum soliton states.

The same GP family also provided a semiclassical
prediction for the long-time variance through
second moments of its Fourier coefficients.
In the large-ring limit, we further derived an
explicit scaling function for these second moments
and an asymptotic prediction for the total variance.
Numerical comparisons supported the GP predictions
for both individual Fourier modes and the total variance.
We also related the yrast transition frequencies
semiclassically to the velocities of the GP solitons
entering the canonical-momentum average.
We found that the weighted frequency width
characterized initial dephasing, whereas the
representative adjacent spacing was associated
with a longer time scale for finite-time averaging.
The numerical collapse of the initial-dephasing
curves and the comparison of the two frequency
scales supported this separation.
We also examined the total finite-window density
variance and compared it with its infinite-time value.

A challenging open problem is to prove the proposed convergence of the
quantum density and its time evolution to the GP average in the weak-coupling
semiclassical limit. This requires asymptotic control of the density form
factors as well as the yrast energies. Establishing the long-time variance
prediction also requires control of the sum over spatial modes.
It would also be useful to extend the present
GP averaging and fluctuation analysis to other
yrast-state weights and determine how the resulting
classical averages affect the localization and lifetime
of quantum soliton profiles.
Related constructions can be found in
\cite{KaminishiMoriMiyashita2020,SarviVianelloCaux2026}.

\section*{Acknowledgements}
The authors thank Tomoyuki Takenawa for useful discussions.
This work grew out of our earlier investigations of
density fluctuations and relaxation times of quantum
dark solitons, which included conjectures on scaling
exponents motivated by exact results in the
Tonks--Girardeau regime.
ChatGPT (GPT-5.6) was used to assist with
research discussions, mathematical and numerical
analyses, and English-language editing.
The authors are responsible for the content
of the manuscript.
This work was partially supported by Grant-in-Aid for Scientific Research (C) 26K06839.

\appendix
\numberwithin{equation}{section}
\section{The canonical-momentum parametrization}\label{app:GP-param-details}

\paragraph{Admissible elliptic parameters.}
At fixed $\Lambda$, \eqref{eq:kappa-def} and \eqref{eq:R-roots-main} give
$R_1=1-4K(K-E)/\Lambda^2$. Both $K$ and $K-E$ are positive and strictly
increasing for $0<m_e<1$, as follows from their integral definitions.
Thus $R_1$ decreases strictly from $1$ to $-\infty$ and has a unique
zero $m_*(\Lambda)$. The condition $R_1\ge0$ is equivalent to
$0\le m_e\le m_*(\Lambda)$. Since $K(m_*)[K(m_*)-E(m_*)]=\Lambda^2/4$,
the endpoint $m_*(\Lambda)$ increases to $1$ as $\Lambda\to\infty$.

\paragraph{Strict monotonicity of $F_\Lambda$.}
Write $m=m_e$ in this proof and let primes denote derivatives with respect
to $m$ at fixed $\Lambda$. Set $P=\Pi(\eta\mid m)$ and $D=\sqrt{R_1R_2R_3}$.
Then \eqref{eq:F-Lambda-main} is $F_\Lambda=\Lambda D(P/(KR_1)-1)/(2\pi)$.
The differentiation identities
\begin{align*}
 K'&=\frac{E-(1-m)K}{2m(1-m)},&
 E'&=\frac{E-K}{2m},\\
 \partial_m\Pi&=\frac{P-E/(1-m)}{2(\eta-m)},&
 \partial_\eta\Pi
 &=\frac{\eta E+(m-\eta)K+(\eta^2-m)P}{2(m-\eta)(\eta-1)\eta}
\end{align*}
give a useful cancellation in the derivative of $DP/(KR_1)$.
Indeed, using $R_2=R_1(1-\eta)$, $R_3=R_1(m-\eta)/m$ and
$\eta=-4mK^2/(\Lambda^2R_1)$, the coefficient of $P$ in $P'$ is
\[
 \frac{1}{2(\eta-m)}
 +\frac{\eta'}2\left(\frac1\eta-\frac1{\eta-1}-\frac1{\eta-m}\right)
 =\frac{K'}K+\frac{R_1'}{R_1}-\frac{D'}{D}.
\]
This cancels the derivative of the prefactor $D/(KR_1)$. The other
terms simplify by $\eta E+(m-\eta)K=mK/R_1$, leaving
\begin{equation}\label{eq:GP-F-after-cancellation-app}
 \frac{2\pi F_\Lambda'}{\Lambda D}
 =\frac{E}{2m(1-m)K R_3}
  -\frac{1}{2R_2R_3}\frac{\eta'}{\eta}
  -\frac12\left(\frac{R_1'}{R_1}+\frac{R_2'}{R_2}+\frac{R_3'}{R_3}\right).
\end{equation}
For clarity, the derivatives obtained from \eqref{eq:R-roots-main} are
\[
 \begin{aligned}
 R_1'&=\frac{2[(K-E)^2-mK^2]}{\Lambda^2m(1-m)},\\
 R_2'&=\frac{2[(K-E)^2+mK(2E-K)]}{\Lambda^2m(1-m)},\\
 R_3'&=\frac{2[E^2-(1-m)K^2]}{\Lambda^2m(1-m)},\qquad
 \frac{\eta'}{\eta}=\frac1m+2\frac{K'}K-\frac{R_1'}{R_1}.
 \end{aligned}
\]
Substituting them into \eqref{eq:GP-F-after-cancellation-app} and putting
the terms over a common denominator gives, with $d=1-E/K$,
\begin{equation*}
 \begin{aligned}
 F_\Lambda'(m)&=\frac{8K^4}{\pi\Lambda^3m(1-m)D}
 \biggl[d(m-d)(1-d)\\
 &\hspace{2.8em}+\frac{K^2}{\Lambda^2}
 \left\{\bigl(m-d^2(3-2d)\bigr)^2+4d^3(1-d)^3\right\}\biggr].
 \end{aligned}
\end{equation*}
Here $d/m$ is the weighted average of $\sin^2\alpha$ on $[0,\pi/2]$
with weight $(1-m\sin^2\alpha)^{-1/2}$. Hence $0<d<m<1$ and
$F_\Lambda'(m)>0$ throughout $0<m<m_*(\Lambda)$.

\paragraph{Endpoints.}
At $m_e=0$, $R=1$ and $I_R=\Lambda$, so $F_\Lambda(0)=0$.
At the other endpoint, the phase integral $I_R$ diverges while
$\sqrt{R_1R_2R_3}$ vanishes. To evaluate their product, substitute
$\alpha=\operatorname{am}(\kappa z\mid m_e)$ in the period integral
$I_R$, use symmetry about $\alpha=\pi/2$, and then set
$\tan\alpha=\sqrt{R_1/R_2}\,t$. This gives
\begin{equation*}
 \sqrt{R_1R_2R_3}\, I_R
 =\frac{2\sqrt{R_3}}{\kappa}
 \int_0^\infty\frac{1}{1+t^2}
 \sqrt{\frac{1+(R_1/R_2)t^2}{1+(1-m_e)(R_1/R_2)t^2}}\,\dd t.
\end{equation*}
As $m_e\uparrow m_*$, the square-root factor tends to one and is bounded
by $(1-m_*)^{-1/2}$. Also $\kappa^2\to R_3(m_*)$. Dominated convergence
therefore gives $\sqrt{R_1R_2R_3}\, I_R\to\pi$. By
\eqref{eq:F-Lambda-main} and \eqref{eq:GP-velocity-winding},
\[
 F_\Lambda=\frac{\sqrt{R_1R_2R_3}}{2\pi}(I_R-\Lambda)
 \longrightarrow\frac12.
\]
This extends $F_\Lambda$ to a continuous bijection from
$[0,m_*(\Lambda)]$ to $[0,1/2]$ and yields the black-soliton limit $s=1/2$ of the
two branches in \eqref{eq:GP-momentum-choices}.

\section{GP soliton limits and Fourier estimates}
\label{app:classical-estimates}

\subsection{Proof of local uniform convergence of the static and moving GP averages}

\noindent\textbf{Uniform bound on $\kappa_\Lambda(s)$.}
For each admissible periodic soliton with canonical momentum \(s\), we write
\[
\kappa=\kappa_\Lambda(s)
:=\frac{2K(m_e(s;\Lambda))}{\Lambda}.
\]
Since $R_1\ge0$, \eqref{eq:R-roots-main} gives
\[
 \kappa^2\le 1+\frac{2\kappa E(m_e)}{\Lambda}.
\]
Combining this with $E(m_e)\le\pi/2$, we further obtain
\begin{equation}\label{eq:app-kappa-bound}
 \kappa^2\le1+\frac{\pi\kappa}{\Lambda}.
\end{equation}
For $\Lambda\ge\pi$, this implies $\kappa^2\le1+\kappa$, and hence
$\kappa_\Lambda(s)<2$ uniformly in $s$. We may therefore take $M=2$
as a uniform upper bound in the estimates below.

\medskip
\noindent\textbf{Convergence of the soliton parameters.}
To prove convergence of the periodic GP soliton profiles at fixed
canonical momentum $s$, we first determine the limits of
$\kappa_\Lambda(s)$ and $m_e(s;\Lambda)$.
For fixed $0<\kappa<1$, let $m_e(\kappa;\Lambda)$ be determined by the
periodicity condition \eqref{eq:decompact-K-condition}.
The limit \eqref{eq:F-infinite-kappa} gives
\[
 F_\Lambda\bigl(m_e(\kappa;\Lambda)\bigr)
 \longrightarrow
 f(\kappa):=\frac{\arcsin\kappa-\kappa\sqrt{1-\kappa^2}}{\pi},
\]
where $f$ is strictly increasing.

For fixed $s\in(0,1/2)$, choose
$0<\kappa_-<f^{-1}(s)<\kappa_+<1$.
Since $f(\kappa_-)<s<f(\kappa_+)$, applying
\eqref{eq:F-infinite-kappa} to the fixed values $\kappa_\pm$ gives,
for sufficiently large $\Lambda$,
\[
 F_\Lambda\bigl(m_e(\kappa_-;\Lambda)\bigr)
 <s<
 F_\Lambda\bigl(m_e(\kappa_+;\Lambda)\bigr).
\]
Since $F_\Lambda(m_e(s;\Lambda))=s$ by \eqref{eq:GP-me-of-s}
and $F_\Lambda$ is strictly increasing
(Appendix~\ref{app:GP-param-details}), we have
\[
 m_e(\kappa_-;\Lambda)
 <m_e(s;\Lambda)
 <m_e(\kappa_+;\Lambda).
\]
Applying the monotonicity of $K(m_e)$, together with
$K(m_e(\kappa_\pm;\Lambda))=\kappa_\pm\Lambda/2$
from \eqref{eq:decompact-K-condition} and
$\kappa_\Lambda(s)=2K(m_e(s;\Lambda))/\Lambda$
from \eqref{eq:kappa-def}, yields
\[
 \kappa_-<\kappa_\Lambda(s)<\kappa_+.
\]
Letting $\kappa_\pm\to f^{-1}(s)$, we obtain
\[
 \kappa_\Lambda(s)\longrightarrow
 f^{-1}(s)=\sqrt{1-\xi(s)^2},
\]
where $\xi(s)$ is the inverse function of the canonical-momentum
relation $s(\xi)$ defined in \eqref{eq:s-xi-infinite}.
Moreover, $K(m_e(s;\Lambda))=\kappa_\Lambda(s)\Lambda/2\to\infty$,
so $m_e(s;\Lambda)\to1$.

By the symmetry $s\leftrightarrow1-s$, both limits extend to
$1/2<s<1$.
\par\medskip
\noindent\textbf{Convergence of individual soliton profiles.}
Having established the limits of the soliton parameters, we now
consider the periodic density
\begin{equation*}
 R_{s,\Lambda}^{\rm per}(y)
 =1+\frac{2\kappa_\Lambda(s)E(m_e(s;\Lambda))}{\Lambda}
 -\kappa_\Lambda(s)^2
 \dn^2\!\left(\kappa_\Lambda(s)y\mid m_e(s;\Lambda)\right).
\end{equation*}
Using $E(m_e)\le\pi/2$, the uniform bound
$\kappa_\Lambda(s)\le M=2$ obtained from
\eqref{eq:app-kappa-bound}, and the nonnegativity of the periodic
density, we obtain
\begin{equation}\label{eq:app-profile-bound}
 0\le R_{s,\Lambda}^{\rm per}(y)
 \le1+\frac{\pi M}{\Lambda}\le3.
\end{equation}
This bound is uniform in $s$ and $y$ and will be used below to justify
convergence of the canonical-momentum average.

For each fixed $s\in(0,1)\setminus\{1/2\}$,
the limits of $\kappa_\Lambda(s)$ and $m_e(s;\Lambda)$
established above, together with
\eqref{eq:decompact-elliptic-limits}, give
\[
 \begin{aligned}
 R_{s,\Lambda}^{\rm per}(y)
 &\longrightarrow 1-(1-\xi(s)^2)
 \sech^2\!\left(\sqrt{1-\xi(s)^2}\,y\right) =R_{\xi(s)}(y).
 \end{aligned}
\]
The convergence is uniform on every bounded $y$-interval.
The Jacobi-function differentiation formulas also imply, for any fixed
$k\ge0$ and $Y>0$,
\[
 \sup_{|y|\le Y}
 \left|\partial_y^kR_{s,\Lambda}^{\rm per}(y)
       -\partial_y^kR_{\xi(s)}(y)\right|
 \longrightarrow0.
\]
Moreover, since $0\le m_e\le1$, $\kappa_\Lambda(s)\le M$, and the
Jacobi functions are bounded in absolute value by $1$ on the real axis,
their differentiation formulas give
\[
 \sup_{\substack{\Lambda\ge\pi\\s\in(0,1)\\y\in\mathbb R}}
 \left|\partial_y^kR_{s,\Lambda}^{\rm per}(y)\right|
 \le C_k<\infty,\qquad k=1,2,\ldots,
\]
where $C_k$ depends only on $k$.

\medskip
\noindent\textbf{Convergence of the static average.}
We now establish the local uniform convergence of the
canonical-momentum average. For $0<\delta,\varepsilon<1/4$, define
\[
 S_{\delta,\varepsilon}
 = [\delta,1/2-\varepsilon]\cup[1/2+\varepsilon,1-\delta].
\]
On each interval of $S_{\delta,\varepsilon}$, $\kappa_\Lambda(s)$
is monotone in $s$ and converges pointwise to the continuous
function $\sqrt{1-\xi(s)^2}$, where $\xi(s)$ is defined
by \eqref{eq:s-xi-infinite}. Hence the convergence is uniform
on $S_{\delta,\varepsilon}$:
\[
 \sup_{s\in S_{\delta,\varepsilon}}
 \left|\kappa_\Lambda(s)-\sqrt{1-\xi(s)^2}\right|
 \longrightarrow0.
\]
Since the limiting function is bounded away from zero on
$S_{\delta,\varepsilon}$, the relation
$K(m_e(s;\Lambda))=\kappa_\Lambda(s)\Lambda/2$ also gives
$m_e(s;\Lambda)\to1$ uniformly there. Together with the
Jacobi-function limits \eqref{eq:decompact-elliptic-limits},
this yields
\begin{equation}\label{eq:app-interior-profile-conv}
 \sup_{\substack{|y|\le Y\\s\in S_{\delta,\varepsilon}}}
 \left|R_{s,\Lambda}^{\rm per}(y)-R_{\xi(s)}(y)\right|
 \longrightarrow0.
\end{equation}
To extend the convergence to the canonical-momentum average over
$s\in[0,1]$, we split the integral into $S_{\delta,\varepsilon}$ and
its complement, whose total measure is $2\delta+2\varepsilon$.
Using the uniform bound \eqref{eq:app-profile-bound} and
$0\le R_\xi\le1$, we obtain
\[
 \begin{aligned}
 \sup_{|y|\le Y}
 \left|R_{\mathrm{av},\Lambda}^{\rm per}(y)-R_{\rm av}(y)\right|
 &\le \sup_{|y|\le Y}
 \int_{S_{\delta,\varepsilon}}
 \left|R_{s,\Lambda}^{\rm per}(y)-R_{\xi(s)}(y)\right|\,\dd s
 +8(\delta+\varepsilon)\\
 &\le \sup_{\substack{s\in S_{\delta,\varepsilon}\\|y|\le Y}}
 \left|R_{s,\Lambda}^{\rm per}(y)-R_{\xi(s)}(y)\right|
 +8(\delta+\varepsilon).
 \end{aligned}
\]
The last inequality uses $|S_{\delta,\varepsilon}|\le1$.
The first term tends to zero by \eqref{eq:app-interior-profile-conv}.
Taking first $\Lambda\to\infty$ and then
$\delta,\varepsilon\downarrow0$ proves the claim for $k=0$.
The same argument applies to derivatives of any fixed order, using
the convergence and uniform derivative bounds established above.
Consequently,
\begin{equation*}
 \sup_{|y|\le Y}
 \left|\partial_y^k R_{\mathrm{av},\Lambda}^{\rm per}(y)
 -\partial_y^kR_{\rm av}(y)\right|\longrightarrow0,
 \qquad k=0,1,2,\ldots.
\end{equation*}
This proves \eqref{eq:classical-profile-decompact-main}.

\medskip
\noindent\textbf{Convergence of the moving average.}
Let \(u_\Lambda(s)\) denote the velocity \(u(s)\) in \(\eqref{eq:GP-velocity-s}\), with its dependence on \(\Lambda\) made explicit. We finally prove the convergence of the time-dependent GP average by combining the convergence of the periodic soliton profiles with the velocity limit
\[
 u_\Lambda(s)\longrightarrow2\xi(s),
 \qquad s\in(0,1)\setminus\{1/2\},
\]
established in Section~\ref{subsec:dyn-infinite-line} using
\eqref{eq:GP-A-Lambda-main} and \eqref{eq:GP-velocity-s}.

For fixed $Y,T>0$, define
\[
 \begin{aligned}
 H_\Lambda(s):=
 \sup_{\substack{|y|\le Y\\|\tau|\le T}}
 \bigl|&R_{s,\Lambda}^{\rm per}(y-u_\Lambda(s)\tau)
       -R_{\xi(s)}(y-2\xi(s)\tau)\bigr|.
 \end{aligned}
\]
The local uniform convergence of the individual soliton profiles,
together with the velocity limit, gives
\[
 H_\Lambda(s)\longrightarrow0
 \qquad\text{for every }s\in(0,1)\setminus\{1/2\}.
\]
Moreover, the uniform density bound \eqref{eq:app-profile-bound} and
$0\le R_\xi\le1$ imply
\[
 0\le H_\Lambda(s)\le4
\]
for all sufficiently large $\Lambda$, uniformly in $s$.
Thus, the dominated convergence theorem applies and yields
\[
 \lim_{\Lambda\to\infty}\int_0^1H_\Lambda(s)\,\dd s
 =\int_0^1\lim_{\Lambda\to\infty}H_\Lambda(s)\,\dd s
 =0.
\]
Consequently,
\[
 \begin{aligned}
 &\sup_{\substack{|y|\le Y\\|\tau|\le T}}
 \left|R_{\mathrm{dyn},\Lambda}^{\rm GP}(y,\tau)
       -R_{\rm dyn,\infty}(y,\tau)\right|
       \le\int_0^1H_\Lambda(s)\,\dd s
 \longrightarrow0.
 \end{aligned}
\]
This proves \eqref{eq:dyn-decompact-uniform}.
The convergence holds on bounded space--time intervals, with
$Y$ and $T$ fixed.

\subsection{Fourier coefficients and weighted convergence}

The nonzero periodic Fourier coefficients underlie both the infinite-line
Fourier limit in Section~\ref{sec:infinite-line-average} and the GP
fluctuation moments in Section~\ref{subsec:fluctuation-gp}.
Using $r=m/\Lambda$ from \eqref{eq:r-healing-main} and the
representation of the GP Fourier coefficients \eqref{eq:gp-csch-fourier-main},
we define, for $r>0$,
\begin{equation}\label{eq:app-b-cont}
 \mathfrak b_\Lambda(s,r):=
 -2\pi^2r\,\csch\!\left(
 \frac{2\pi r K(1-m_e(s;\Lambda))}{\kappa_\Lambda(s)}\right).
\end{equation}
We extend $\mathfrak b_\Lambda(s,r)$ continuously to $r=0$.
For $m\ge1$, this agrees with the periodic Fourier coefficient,
$\mathfrak b_\Lambda(s,m/\Lambda)=\mathfrak b_{m,\Lambda}^{\rm GP}(s)$.
The continuous extension at $r=0$ should not be confused with the true
zero Fourier coefficient of the finite-ring density deviation, which vanishes.

Define the finite-ring first and second canonical-momentum moments by
\begin{equation}\label{eq:app-BPhi-cont}
 B_\Lambda(r):=\int_0^1\mathfrak b_\Lambda(s,r)\,\dd s,
 \qquad
 \Phi_\Lambda(r):=2\int_0^1
 |\mathfrak b_\Lambda(s,r)|^2\,\dd s.
\end{equation}
At $r=m/\Lambda$, $B_\Lambda(r)$ is the first moment defined in
\eqref{eq:A-first-moment}, while $\Lambda^{-2}\Phi_\Lambda(r)$ is the
second moment defined in \eqref{eq:A-second-moment}. Their infinite-line
counterparts are $B_{\rm av}(r)$ and $\Phi(r)$, defined in
\eqref{eq:Bav-integral-main} and \eqref{eq:Phi-def-main}, respectively.
Theorem~\ref{thm:app-weighted} below establishes the convergence
of $B_\Lambda(r)$ to $B_{\rm av}(r)$ and of $\Phi_\Lambda(r)$ to
$\Phi(r)$.

\begin{lemma}\label{lem:app-majorant}
There exist constants $C,c_0>0$, independent of $s,r,$ and $\Lambda$,
such that, for all sufficiently large $\Lambda$ and $r\ge0$,
\begin{equation}\label{eq:app-majorant}
 |\mathfrak b_\Lambda(s,r)|
 +|\partial_r\mathfrak b_\Lambda(s,r)|
 \le Ce^{-c_0r}.
\end{equation}
Consequently,
\[
\begin{aligned}
 |B_\Lambda(r)|+|B_\Lambda'(r)|&\le Ce^{-c_0r},\\
 |\Phi_\Lambda(r)|+|\Phi_\Lambda'(r)|&\le C'e^{-2c_0r},
\end{aligned}
\]
where $C'>0$ is independent of $\Lambda$.
\end{lemma}
\begin{proof}
As shown in Appendix B.1 using \eqref{eq:app-kappa-bound},
$\kappa_\Lambda(s)\le M=2$ for $\Lambda\ge\pi$.
Since $K(q)\ge\pi/2$ for $0\le q<1$,
\[
\begin{aligned}
 a_\Lambda(s)
 &:=\frac{2\pi K(1-m_e(s;\Lambda))}{\kappa_\Lambda(s)}
 \ge\frac{\pi^2}{\kappa_\Lambda(s)}
 \ge\frac{\pi^2}{M}=:a_0>0.
\end{aligned}
\]
Using $\csch x=1/\sinh x$ in \eqref{eq:app-b-cont} and differentiating
with respect to $r$, we obtain, for $r>0$,
\[
\begin{aligned}
 |\mathfrak b_\Lambda(s,r)|
 &=\frac{2\pi^2}{a_\Lambda(s)}
 \frac{a_\Lambda(s)r}{\sinh(a_\Lambda(s)r)},\\[2mm]
 |\partial_r\mathfrak b_\Lambda(s,r)|
 &=2\pi^2\left|
 \frac{\sinh(a_\Lambda(s)r)-a_\Lambda(s)r\cosh(a_\Lambda(s)r)}
 {\sinh^2(a_\Lambda(s)r)}\right|.
\end{aligned}
\]
To bound these expressions, observe that
\[
 \frac{x}{\sinh x}=1+O(x^2),\qquad
 \frac{\sinh x-x\cosh x}{\sinh^2x}=O(x)
 \qquad(x\to0),
\]
while both expressions are $O((1+x)e^{-x})$ as $x\to\infty$.
Consequently, for some $C_1>0$,
\[
 \frac{x}{\sinh x}
 +\left|\frac{\sinh x-x\cosh x}{\sinh^2x}\right|
 \le C_1 e^{-x/2},\qquad x>0.
\]
Applying this inequality with $x=a_\Lambda(s)r$, and using
$a_\Lambda(s)\ge a_0$, gives
\[
\begin{aligned}
 |\mathfrak b_\Lambda(s,r)|
 &\le\frac{2\pi^2C_1}{a_\Lambda(s)}e^{-a_\Lambda(s)r/2}
 \le\frac{2\pi^2C_1}{a_0}e^{-a_0r/2},\\[2mm]
 |\partial_r\mathfrak b_\Lambda(s,r)|
 &\le2\pi^2C_1e^{-a_\Lambda(s)r/2}
 \le2\pi^2C_1e^{-a_0r/2}.
\end{aligned}
\]
These bounds extend to $r=0$ by continuity, proving
\eqref{eq:app-majorant} with $c_0=a_0/2$.

The bounds for $B_\Lambda$, $\Phi_\Lambda$, and their first derivatives
follow by differentiating the $s$-integrals in
\eqref{eq:app-BPhi-cont}, justified by \eqref{eq:app-majorant}.
\end{proof}

\begin{theorem}\label{thm:app-weighted}
For every $R<\infty$,
\begin{equation}\label{eq:app-local-moment-conv}
 \begin{aligned}
 \sup_{0\le r\le R}|B_\Lambda(r)-B_{\rm av}(r)|&\longrightarrow0,\\
 \sup_{0\le r\le R}|\Phi_\Lambda(r)-\Phi(r)|&\longrightarrow0.
 \end{aligned}
\end{equation}
For every fixed integer $k\ge0$,
\begin{align}
 \int_0^\infty r^k|B_\Lambda(r)-B_{\rm av}(r)|\,\dd r&\longrightarrow0,
 \label{eq:app-B-L1}\\
 \int_0^\infty r^k|\Phi_\Lambda(r)-\Phi(r)|\,\dd r&\longrightarrow0.
 \label{eq:app-Phi-L1}
\end{align}
\end{theorem}
\begin{proof}
For each fixed $s\in(0,1)\setminus\{1/2\}$, the parameter limits
established in Appendix B.1 give
\[
 \kappa_\Lambda(s)\longrightarrow\sqrt{1-\xi(s)^2},\qquad
 K(1-m_e(s;\Lambda))\longrightarrow\frac\pi2.
\]
Consequently,
\[
 a_\Lambda(s)\longrightarrow
 \frac{\pi^2}{\sqrt{1-\xi(s)^2}}>0.
\]
By \eqref{eq:app-b-cont} and \eqref{eq:Bxi-main}, respectively,
\[
\begin{aligned}
 \mathfrak b_\Lambda(s,r)
 &=-2\pi^2r\,\csch\!\left(a_\Lambda(s)r\right),\\
 B_{\xi(s)}(r)
 &=-2\pi^2r\,\csch\!\left(
   \frac{\pi^2r}{\sqrt{1-\xi(s)^2}}\right).
\end{aligned}
\]
Both expressions are extended continuously to $r=0$. Since
$(a,r)\mapsto-2\pi^2r\csch(ar)$ is uniformly continuous
on compact subsets of $(0,\infty)\times[0,\infty)$, the convergence
of $a_\Lambda(s)$ to a positive limit implies, for every fixed
$s\in(0,1)\setminus\{1/2\}$,
\[
 \sup_{0\le r\le R}
 |\mathfrak b_\Lambda(s,r)-B_{\xi(s)}(r)|\longrightarrow0.
\]
Using the definitions of $B_\Lambda$ and $B_{\rm av}$
in \eqref{eq:app-BPhi-cont} and \eqref{eq:Bav-integral-main}, respectively,
we have
\[
\begin{aligned}
 &\sup_{0\le r\le R}|B_\Lambda(r)-B_{\rm av}(r)|\le\int_0^1
 \sup_{0\le r\le R}|\mathfrak b_\Lambda(s,r)-B_{\xi(s)}(r)|\,\dd s.
\end{aligned}
\]
The integrand on the right converges to zero for almost every $s$
and is bounded by $2C$, independently of $\Lambda$, by
Lemma~\ref{lem:app-majorant}. The dominated convergence theorem therefore gives
\[
\begin{aligned}
 &\lim_{\Lambda\to\infty}\int_0^1
 \sup_{0\le r\le R}|\mathfrak b_\Lambda(s,r)-B_{\xi(s)}(r)|\,\dd s \\
 =&\int_0^1\lim_{\Lambda\to\infty}
 \sup_{0\le r\le R}|\mathfrak b_\Lambda(s,r)-B_{\xi(s)}(r)|\,\dd s
 =0.
\end{aligned}
\]
Using the definitions of $\Phi_\Lambda$ and $\Phi$ in
\eqref{eq:app-BPhi-cont} and \eqref{eq:Phi-def-main}, the same argument
applies to the squared coefficients, since
\[
 \bigl||z|^2-|w|^2\bigr|\le(|z|+|w|)|z-w|.
\]
This proves both limits in \eqref{eq:app-local-moment-conv}.

We next prove the weighted convergence. The exponential bounds for
$B_\Lambda(r)$ and $\Phi_\Lambda(r)$ in Lemma~\ref{lem:app-majorant}
also hold for their limiting functions $B_{\rm av}(r)$ and $\Phi(r)$,
respectively, by \eqref{eq:app-local-moment-conv}. Hence
\[
\begin{aligned}
 r^k|B_\Lambda(r)-B_{\rm av}(r)|
 &\le2Cr^ke^{-c_0r},\\
 r^k|\Phi_\Lambda(r)-\Phi(r)|
 &\le2C'r^ke^{-2c_0r}.
\end{aligned}
\]
Both right-hand sides are integrable on $[0,\infty)$.
Since the left-hand sides converge pointwise to zero by
\eqref{eq:app-local-moment-conv}, dominated convergence in $r$
yields \eqref{eq:app-B-L1} and \eqref{eq:app-Phi-L1}.

\end{proof}

\subsection{GP second moments and limiting regimes}
\label{app:fluctuation-second-moment}

\medskip
\noindent\textbf{Fourier second moments.}
We start from \eqref{eq:Phi-integral-main}.
By symmetry and the substitution
$\sin\alpha=\sech t$, we have
\begin{align*}
\Phi(r)
&=16\pi^3r^2\int_0^\pi
 \sin^2\alpha\,
 \csch^2\left(\frac{\pi^2r}{\sin\alpha}\right)
 \,\dd\alpha\\
&=32\pi^3r^2\int_0^\infty
 \sech^3t\,
 \csch^2(\pi^2r\cosh t)\,\dd t.
\end{align*}
Using the expansion
$\csch^2z=4\sum_{n\ge1}n e^{-2nz}$
and the definition \eqref{eq:Ki-def-main},
we obtain
\begin{equation*}
 \Phi(r)=128\pi^3r^2
 \sum_{n=1}^\infty
 n\,\operatorname{Ki}_3(2n\pi^2r),
 \qquad r>0.
\end{equation*}
Since $x^2\csch^2x\le1$ for $x\ge0$, the limiting integrand of
\eqref{eq:Phi-integral-main} is dominated by an
integrable function on $[0,\pi]$.  At $r=0$ its value is
$16\sin^4\alpha/\pi$, so
\[
 \Phi(0)=\frac{16}{\pi}\int_0^\pi\sin^4\alpha\,\dd\alpha=6.
\]
Using the explicit transform
\eqref{eq:Bxi-main} and the elementary integral
$\int_0^\infty x^2\csch^2x\,\dd x=\pi^2/6$, we obtain
\[
 \int_0^\infty|B_\xi(r)|^2\,\dd r
 =\frac23(1-\xi^2)^{3/2}.
\]
It follows from \eqref{eq:measure-change-main} that
\[
 I_1=\int_0^\infty\Phi(r)\,\dd r
 =\frac8{3\pi}\int_{-1}^1(1-\xi^2)^2\,\dd\xi
 =\frac{128}{45\pi}.
\]
These are the two constants given by
\eqref{eq:Phi-constants-main}.

\medskip
\noindent\textbf{From Fourier integrals to discrete mode sums.}
To pass from the continuum variable $r$ back to the discrete Fourier
modes $r=m/\Lambda$, we use the following elementary estimate.
\begin{lemma}\label{lem:app-riemann}
If $f:[0,\infty)\to\mathbb C$ is absolutely continuous on bounded
intervals, with $f,f'\in L^1(0,\infty)$, then for every $h>0$,
\begin{equation}\label{eq:app-riemann}
 \left|h\sum_{m=1}^\infty f(mh)
 -\int_0^\infty f(r)\,\dd r\right|
 \le h\int_0^\infty|f'(r)|\,\dd r.
\end{equation}
\end{lemma}
\begin{proof}
Dividing $[0,\infty)$ into intervals of length $h$ and using
$f(mh)-f(r)=\int_r^{mh}f'(u)\,\dd u$, we obtain
\[
\begin{aligned}
 &\left|h\sum_{m=1}^\infty f(mh)
   -\int_0^\infty f(r)\,\dd r\right|
 \le\sum_{m=1}^\infty
   \int_{(m-1)h}^{mh}|f(mh)-f(r)|\,\dd r\\
 \le&\sum_{m=1}^\infty
   \int_{(m-1)h}^{mh}\int_r^{mh}|f'(u)|\,\dd u\,\dd r
 \le h\int_0^\infty|f'(u)|\,\dd u.
\end{aligned}
\]
This proves \eqref{eq:app-riemann}.
\end{proof}

The exponential bounds in Lemma~\ref{lem:app-majorant} give
a uniform bound on $\int_0^\infty|\Phi_\Lambda'(r)|\,\dd r$.
Applying Lemma~\ref{lem:app-riemann} with $h=\Lambda^{-1}$
to $\Phi_\Lambda$, we obtain
\[
 \left|\frac1\Lambda\sum_{m\ge1}\Phi_\Lambda(m/\Lambda)
 -\int_0^\infty\Phi_\Lambda(r)\,\dd r\right|
 \le\frac1\Lambda\int_0^\infty|\Phi_\Lambda'(r)|\,\dd r
 =O(\Lambda^{-1}).
\]
Together with the $L^1$-convergence \eqref{eq:app-Phi-L1}, this gives
\[
 \frac1\Lambda\sum_{m\ge1}\Phi_\Lambda(m/\Lambda)
 \longrightarrow I_1.
\]
Since $G_{m,\Lambda}^{\rm GP}=\Lambda^{-2}\Phi_\Lambda(m/\Lambda)$,
this is equivalently
\begin{equation*}
 \Lambda\sum_{m\ge1}G_{m,\Lambda}^{\rm GP}\longrightarrow I_1.
\end{equation*}
For $R\ge0$, the cumulative fraction of the finite-ring GP mode
weights satisfies
\begin{equation}\label{eq:app-cumulative-GP}
 \frac{\displaystyle\sum_{1\le m\le\lfloor\Lambda R\rfloor}
       G_{m,\Lambda}^{\rm GP}}
      {\displaystyle\sum_{m\ge1}G_{m,\Lambda}^{\rm GP}}
 =\frac{\displaystyle\Lambda^{-1}
       \sum_{1\le m\le\lfloor\Lambda R\rfloor}
       \Phi_\Lambda(m/\Lambda)}
      {\displaystyle\Lambda^{-1}
       \sum_{m\ge1}\Phi_\Lambda(m/\Lambda)}
 \longrightarrow\frac1{I_1}\int_0^R\Phi(r)\,\dd r.
\end{equation}
For each fixed $R\ge0$, the convergence follows from the
Riemann-sum estimate in Lemma~\ref{lem:app-riemann} and the
$L^1$-convergence \eqref{eq:app-Phi-L1} with $k=0$.
Since
$I_1^{-1}\int_0^{0.216}\Phi(r)\,\dd r\simeq0.9$,
the modes $m\lesssim0.216\Lambda$ carry approximately $90\%$ of the
total finite-ring GP mode weight at large $\Lambda$, as quoted in
Section~\ref{subsec:fluctuation-gp}.

\medskip
\noindent\textbf{Ultraweak coupling and the
small-$\Lambda$ limit.}
Recall from \eqref{eq:prelim-Lambda} that
$\Lambda=N\sqrt{\gamma}$, where
$\gamma=c/\bar n$ and $\bar n=N/L$.
Thus,
$
 Lc=N\gamma=N^{-1} \Lambda^2.
$
The ultraweak-coupling condition $Lc\ll1$
should therefore be distinguished from
$\Lambda\ll1$. In particular, every
fixed-$\Lambda$ sequence eventually satisfies
$Lc\ll1$ as $N\to\infty$.
The condition is also compatible with
$\Lambda\to\infty$; for example,
$\gamma=N^{-3/2}$ gives
$Lc=N^{-1/2}$ and $\Lambda=N^{1/4}$.

We next examine the limit $c\downarrow0$
at fixed $N$ and $L$, taking the infinite-time
average before the coupling tends to zero.
The one-hole yrast state $|P_p,N\rangle$
 \eqref{eq:onehole-PE} approaches
the free-boson Fock state with $N-p$ particles
of momentum zero and $p$ particles of
momentum $2\pi/L$.
For the density form factors
$F_{p,q}=\langle P_q,N|
\hat\rho(0)|P_p,N\rangle$
defined in \eqref{eq:Fpq-def}, we obtain
\[
 \lim_{c\downarrow0}|F_{p,p+1}|^2
 =\frac{(N-p)(p+1)}{L^2},
 \qquad p=0,\ldots,N-2,
\]
whereas $F_{p,p+m}\to0$ for $m\ge2$.

To examine the transition frequencies
$\omega_{p,m,N}=E_{p+m}-E_p$
defined in \eqref{eq:dyn-transition-frequency},
we expand the yrast energies \eqref{eq:onehole-PE} at small $c$:
\[
 E_p=\left(\frac{2\pi}{L}\right)^2p
 +\frac{c}{L}
 \bigl[N(N-1)+2p(N-p)\bigr]+O(c^2).
\]
Consequently,
\[
 \omega_{p+1,m,N}-\omega_{p,m,N}
 =-\frac{4cm}{L}+O(c^2),
 \qquad 0\le p\le N-m-2.
\]
Thus, at fixed $N,L$, the transition
frequencies within each mode are nonzero
and pairwise distinct for sufficiently
small positive $c$.

Using the variance formulas
\eqref{eq:fluctuation-mode-sum} and
\eqref{eq:Vinfty-nondegenerate-main},
together with $\bar n=N/L$, we obtain
\[
\begin{aligned}
 \lim_{c\downarrow0}NV_{\rm tot,N}(\infty)
 &=\frac{2}{N^3}
   \sum_{p=0}^{N-2}(N-p)(p+1)
 =\frac{(N-1)(N+4)}{3N^2}.
\end{aligned}
\]
Taking the large-$N$ limit then gives
\[
 \lim_{N\to\infty}\lim_{c\downarrow0}
 NV_{\rm tot,N}(\infty)=\frac13,
\]
where the inner limit is taken at fixed $N,L$.

We next compare this result with the small-$\Lambda$
limit of the periodic GP second moments
\eqref{eq:A-second-moment}.
The limit can be derived directly from the
elliptic-function representation of the periodic
GP density.
Recall \eqref{eq:R-sn-main}, \eqref{eq:kappa-def},
and \eqref{eq:R-roots-main}.
Since the condition $R_1\ge0$ requires
$m_e=O(\Lambda^2)$ as $\Lambda\to0$, we set
\[
 \beta_\Lambda:=\frac{\pi^2m_e}{2\Lambda^2}.
\]
Using
\[
 K(m_e)=\frac{\pi}{2}
 \left(1+\frac{m_e}{4}+O(m_e^2)\right),
 \qquad
 E(m_e)=\frac{\pi}{2}
 \left(1-\frac{m_e}{4}+O(m_e^2)\right),
\]
we obtain
\begin{equation} \label{eq:expansionapp}
    R_1=1-\beta_\Lambda+O(\Lambda^2),
 \qquad
 R_2=1+\beta_\Lambda+O(\Lambda^2),
 \qquad
 R_3=\frac{\pi^2}{\Lambda^2}+O(1).
\end{equation}
The relation
\eqref{eq:F-Lambda-main} gives, for
$\beta_\Lambda\to\beta$ with $0\le\beta<1$,
\[
 F_\Lambda(m_e)\longrightarrow
 \frac{1-\sqrt{1-\beta^2}}{2}.
\]
Since $F_\Lambda(m_e(s;\Lambda))
=\min\{s,1-s\}$ by \eqref{eq:GP-me-of-s},
we obtain
\[
 \beta=2\sqrt{s(1-s)}.
\]
The black-soliton value $s=1/2$ follows
by continuity.

Using $\kappa\Lambda\to\pi$ and
$\sn(v\mid m_e)\to\sin v$ as
$m_e\to0$, we therefore find
\begin{equation}\label{eq:GP-small-Lambda-profile}
 R_{s,\Lambda}^{\rm per}(\Lambda z)
 \longrightarrow
 1-2\sqrt{s(1-s)}\cos(2\pi z).
\end{equation}
In particular, the Fourier coefficients
defined in \eqref{eq:gp-fourier-integral-main}
satisfy
\[
 \frac{\mathfrak b_{1,\Lambda}^{\rm GP}(s)}
 {\Lambda}\longrightarrow-\sqrt{s(1-s)},
 \qquad
 \frac{\mathfrak b_{m,\Lambda}^{\rm GP}(s)}
 {\Lambda}\longrightarrow0
 \quad(m\ge2).
\]
To establish the small-$\Lambda$ limit of
$S_{\rm GP}(\Lambda)$, we apply
Parseval's identity to the Fourier expansion
\eqref{eq:gp-scaled-fourier-series-main}.
Together with the definitions
\eqref{eq:A-second-moment} and
\eqref{eq:GP-total-variance}, this gives
\begin{equation}\label{eq:GP-total-variance-parseval}
 S_{\rm GP}(\Lambda)
 =
 \int_0^1\!\dd s
 \int_{-1/2}^{1/2}
 \left[R_{s,\Lambda}^{\rm per}(\Lambda z)-1\right]^2
 \,\dd z.
\end{equation}
The expansions \eqref{eq:expansionapp} and $R_1\ge0$ imply
$\beta_\Lambda\le1+O(\Lambda^2)$, uniformly in $s$.
Hence, by \eqref{eq:R-sn-main},
\[
 0\le R_{s,\Lambda}^{\rm per}(\Lambda z)
 \le R_2\le2+O(\Lambda^2),
\]
uniformly in $s\in[0,1]$ and $z\in[-1/2,1/2]$.
Applying the dominated convergence theorem to
\eqref{eq:GP-total-variance-parseval},
using the profile limit
\eqref{eq:GP-small-Lambda-profile}, we obtain
\begin{equation*}
\begin{aligned}
 \lim_{\Lambda\to0}S_{\rm GP}(\Lambda)
 &=
 \int_0^1\!\dd s
 \int_{-1/2}^{1/2}
 4s(1-s)\cos^2(2\pi z)\,\dd z\\
 &=
 2\int_0^1s(1-s)\,\dd s
 =\frac13.
\end{aligned}
\end{equation*}
This agrees with the quantum result obtained above.

The order of limits is important.
At $c=0$, the transition frequencies
within each mode are degenerate, so
taking the free-particle limit before
the infinite-time average retains
interference terms that are absent
when $c\downarrow0$ is taken afterward.

\FloatBarrier

\end{document}